\documentstyle[12pt]{article}

\catcode`\@=11
\def\numberbysection{\@addtoreset{equation}{section}
        \def\theequation{\thesection.\arabic{equation}}}
\def\beq{\begin{equation}}
\def\eeq{\end{equation}}
\def\half{\frac{1}{2}}
\numberbysection
\begin{document}
\begin{titlepage}
\begin{center}
\hfill DFF 238 / 11 / 95  \\
\vskip 1.in
{\Large
\bf Solving The N-Body Problem in (2+1)-Gravity
\footnote{Work supported in part by M.U.R.S.T., Italy.
\\
DFF 238 / 11 / 95 \\
November 1995 } }
\vskip 0.2in
A. Bellini \\
[.2in] {\em Dipartimento di Fisica, Universit\`a di Firenze, Italy,}
\vskip 0.2in
M. Ciafaloni and P. Valtancoli  \\
\vskip 0.2in {\em Dipartimento di Fisica, Universit\`a di Firenze \\
and INFN, Sezione di Firenze, Italy} \\
\end{center}
\vskip .5in
\begin{abstract}
We construct a non-perturbative, single-valued solution for the metric and the
motion of $N$ interacting particles in $2+1$-Gravity. The solution is explicit
for two particles with any speed and for any number of particles with small
speed. It is based on a mapping from multivalued Minkowskian coordinates to
single-valued ones, which solves the non-abelian monodromies due to particles'
momenta. The two and three-body cases are treated in detail.
\end{abstract}
\vfill
\end{titlepage}
\pagenumbering{arabic}
%
%
%
%

\section{Introduction}

We address, in this paper, a classic issue in gravitation theory \cite{a1},
namely the one of finding the self-consistent metric and the corresponding
motion of $N$ interacting particles. This problem turns out to be solvable
in $2+1$ dimensions \cite{a2}-\cite{a11}, and the solution that we find
\cite{a12} shows several nontrivial features.

Firstly, our solution is regular, i.e., metric and coordinates are
single-valued, or in other words, the metric is singular only at the particle
sites. This is to be contrasted with the spurious singularities, found in
previous studies \cite{a3},\cite{a7} by using the existence of locally
Minkowskian coordinates and / or the relation of ($2+1$) gravity to
Chern-Simons theory \cite{a8},\cite{a9}. In fact in ($2+1$) dimensions the
space is flat outside the ( pointlike ) sources, but the corresponding
Minkowskian coordinates are not single-valued, due to the localized curvature
at the particle sites. Therefore, in such solutions \cite{a10},\cite{a11}, the
metric has spurious singularity tails departing from each particle.

Secondly, we are able to treat particles moving with arbitrary speed and
with arbitrary masses, in some "physical" range consistent with an open
universe. Thus we generalize in a nontrivial way the well known \cite{a2},
\cite{a3} conformal metric of the static limit, which in $2+1$ dimensions is a
rather simple one, due to the lack of a Newtonian force.

Finally, the classical phase space emerging in the explicit form of our metric
shows several links with conformal and Liouville theories, which suggest a
way towards an $N$-body quantum mechanics \cite{a5}, \cite{a6},
\cite{a9} and perhaps a full quantum theory.

A basic reason why the $N$-body problem is solvable in the extended sense
just explained, is that there is no graviton radiation in $2+1$ dimensions.
In fact, physical tensor waves (unlike photons) are not possible with only one
transverse space dimension. Therefore, the gravitational degrees of freedom are
longitudinal, and can propagate instantaneously in the gauge of Coulomb type
\cite{a13}-\cite{a16} that we have proposed in I.

Our method of solution exploits both the existence of multivalued Minkowskian
coordinates and the instantaneous propagation to construct a mapping to
single-valued coordinates with nontrivial metric. In a generalized conformal
gauge, such mapping is based on holomorphic and antiholomorphic
representations of the Minkowskian monodromies, in which the analyticity
properties of the mapping function are a consequence of the instantaneous
propagation.

The Minkowskian monodromies are provided by the Deser-Jackiw and 't-Hooft
\cite{a3} (DJH) "matching conditions" and define the particle sources.
Together with
proper boundary conditions at the particle singularities and at space
infinity they determine both the metric and the motion, up to some residual
gauge freedom, which allows one arbitrary trajectory and one scale parameter.

Since the DJH matching conditions involve Lorentz ( or Poincar\`e )
transformations which leave the particle's Minkowskian momentum invariant,
they form in general a non-abelian group, depending on the $N$ particles'
momenta, with a quite complicated algebra. Nevertheless, it is possible
to construct a spin $\half$ ( or projective ) representation of such
monodromies in terms of independent solutions of a second order differential
equation with Fuchsian singularities \cite{a17} ( the "Riemann-Hilbert"
problem \cite{a18}).

The solution for the mapping function is here explicitly given in the two-body
case, in terms of proper hypergeometric functions, and is given for $N \ge 3$
also, but only in the quasi-static limit, i.e., to first nontrivial order in
the velocities.

Given the mapping function, all components of the metric and all the motion
parameters are provided in terms of quadratures.

We have already given in I an account of our method and of the main features
of our solution. Similar ideas have been, later on, discussed by Welling
\cite{a19}.
The purpose of the present paper is to describe the method and the interesting
features of the two-body problem in full detail, and to investigate the novel
features arising for $N \ge 3$, by describing the explicit solutions in the
quasi-static limit ( not to be confused with the static one ) and some other
simple $N$-body example.

An interesting feature arising for $N \ge 3$ is that the mapping function is
determined not only by the particle singularities, but also by some "apparent
singularities" \cite{a18} which have an invariant meaning ( because they occur
in the Schwarzian derivative \cite{a20} of the mapping function ) but have,
nevertheless,
trivial monodromies. Such apparent singularities carry accessory parameters
which are needed to match the non-abelian particle monodromies, including
the one at space infinity. Here they are studied in detail in the quasi-static
case.

The contents of the paper are as follows. In the introductory Section 2 we
review the known singular solutions and, following I, we define the mapping
problem from Minkowskian to single valued coordinates in our conformal
Coulomb gauge. In Sec. 3 we treat, following I, the two-body problem, and
in particular, the solution for the mapping function, the metric, and the
motion. We also set up the conditions for the mapping to be non-singular, and
to avoid closed timelike curves \cite{a20}. In Sec. 4 we treat the many-body
case,
the apparent singularities, and the explicit solution in the quasi-static
limit. In Sec. 5 we describe in more detail the three-body case, its decoupling
properties, and an interesting $N$-body case characterized by a symmetric
particle configuration. We discuss our results in the conclusive Section 6,
where we also give a few suggestions for the left-over problems. Some details
of the quadratures leading to the motion parameters are given in the Appendix.

\section{ Minkowskian vs single-valued coordinates}

{\large \bf 2.1 General features }${}^{[2-11]}$

A basic property of classical $(2+1)$-Gravity is that the Riemann tensor is
proportional to the Einstein tensor, and thus to the energy-momentum tensor.
More precisely, due to the existence of the invariant $\epsilon$-symbol, we
have

\beq R_{\mu\nu}^{\alpha\beta} \ = \ - \ \epsilon_{\mu\nu\lambda} \
\epsilon^{\alpha\beta\gamma} \ T^\lambda_\gamma , \label{b1}\eeq
where, for pointlike particles, the energy-momentum density is a sum of
delta-functions at the particle sites $x^\mu \ = \ \xi^\mu_i ( \tau )$:
\beq \sqrt{g} \ T^{\mu\nu} \ = \ \sum_{i=1}^N m_i \int d\tau_i \frac{d\xi^\mu}{
d\tau_i} \frac{d\xi^\nu}{d\tau_i} \delta^{(3)} (x-\xi_i(\tau)).
\label{b2}\eeq

Therefore, the space is flat everywhere, except at the particles sites, in
which a singular curvature exists, related to the particles' momenta. This
means
that local Minkowskian coordinates can be extended all around the particles,
but are in general multivalued, i.e., carry non-trivial monodromy
transformations for parallel transport in a closed loop around each particle
site.

The simplest example is for one ( spinless ) particle at rest in the origin.
The space is flat everywhere else, but the loop integral of the connection
is nontrivial, i.e., at a given time,

\beq \oint_{C_0} \ dx^\mu \ {(\Gamma_\mu)}_{\alpha\beta} \ = \
\epsilon_{\alpha\beta 0} \ m , \  \ \  ( 8 \pi G = 1 )
\label{b3}\eeq
where $m$ is the particle mass and $C_0$ encircles the origin. This situation
admits several descriptions, according to the coordinate choice. In Minkowskian
coordinates $X^a \equiv ( T / Z / \bar{Z} )$ the line element is trivial

\beq ds^2 \ = \ \eta_{ab} dX^a dX^b \ = \ dT^2 - {|dZ|}^2, \ \ \ Z \ = \
X + i Y,  \label{b4}\eeq
but there is a cut-out sector, or a branch cut, corresponding to a deficit
angle $m$ \cite{a2},\cite{a3} :

\beq |arg Z| < \pi \alpha, \ \ \  \alpha \equiv 1 - \frac{m}{2\pi}
\label{b5}\eeq
so that values of the $Z$ coordinate above and below the cut are related by

\beq Z_{II} \ = \ e^{-im} Z_I. \label{b6}\eeq

The connection is localized on this cut, so that (\ref{b3}) is satisfied.

On the other hand, the branch point can be eliminated by a coordinate
transformation to single-valued variables $x^\mu \equiv ( t / z / \bar{z})$,
defined by

\beq Z \ = \ z^\alpha \ , \ T \ = \ t \label{b7}\eeq
in such a way that, for $z \rightarrow e^{2\pi i } z$, $ Z \rightarrow e^{-im}
Z $, as required by (\ref{b6}). The corresponding line element is now
nontrivial

\beq ds^2 \ = \ dt^2 - \alpha^2 {|z|^{-\frac{m}{\pi}}} {|dz|^2},
\label{b8}\eeq
yielding the conformal gauge metric, for which the connection is isotropically
distributed around the particle.

The scale change $\rho \ = \ |z|^\alpha$ brings (\ref{b8}) to the conical
gauge form

\beq ds^2 \ = \ dt^2 - d\rho^2 - \alpha^2 \rho^2 d\theta^2 , \ \ \
( 0 \le \theta < 2\pi ) , \label{b9}\eeq
where the well-known conical geometry is transparent.

The discontinuity relation (\ref{b6}) is called the DJH matching condition
\cite{a3}. It is
just a rotation for a particle at rest in the origin. If the particle moves and
is located at $X \ = \ X_1$, the monodromy (\ref{b6}) is boosted to a Poincare'
transformation (Fig. 1 ):

\beq X_{II} - X_1 \ = \ L ( P_1 ) ( X_I - X_1 ), \label{b10}\eeq
where the $O(2,1)$ matrix

$$ L( P_1 ) \ = \ exp ( - i J_a \cdot P_1^a ) \ = \ P exp ( - \oint_{C(X_1)}
\omega_\mu dx^\mu ),  $$
\beq {( \ i \ J_a)}_{bc} \ = \ \epsilon_{abc} , \label{b11}\eeq
is the holonomy of the ( spin ) connection, related to the particle's
Minkowskian momentum $P_1^a$, constant of motion. Correspondingly, the
particles
carry string singularities, or tails , which are needed to yield a precise
determination of the $X^a$'s.

While eliminating the branch cut (\ref{b6}) was trivial, eliminating all tails
is
in general difficult, for the following reasons:

(i) If there are at least two particles, with a relative speed, the problem is
non abelian, i.e. the monodromies do not commute

\beq [ L(P_1) , L(P_2) ] \neq 0 \label{b12}\eeq
and therefore cannot be brought together to the form of a phase transformation.

(ii) The rest frame complex planes \#1, \#2 , etc... are inequivalent, i.e. ,
they are related in general by Lorentz transformations which mix space and
time,
and have a complicated time dependence in a coordinate frame in which, say,
tail \#1 is eliminated.

Fortunately, there is a second simplifying feature in $2+1$ dimensions, i.e.,
there are no transverse gravitons. For a given wave vector, there is only one
transverse dimension, which cannot accomodate tensor waves. As a consequence,
there are only longitudinal degrees of freedom, which can propagate
instantaneously, in a properly chosen gauge.

Indeed, here we shall use the conformal gauge of Coulomb type that we have
proposed in I, which provides an instantaneous propagation. This gauge allows
to deal with all monodromies at a given time, in the same complex plane, and
allows to treat the tails as true branch cuts of analytic functions.

{\large \bf 2.2 Singular solutions  }

In order to make the above reasoning more precise, let us recall the class of
singular solutions that were found \cite{a10}, \cite{a11} in the first order
formalism, which exhibits the known relation to a Chern-Simons Poincare' gauge
theory. By defining dreiben and spin connection in the usual way

$$ g_{\mu\nu} \ \equiv \ E^a_\mu E^b_\nu \eta_{ab} , $$
\beq \Gamma_{\lambda,\mu \nu} \ = \ E^a_{\lambda} \ {( \partial_\mu +
\omega_\mu )}_{ab} \ E^b_\nu , \label{b13}\eeq

such solutions turn out to be additive with the particles, provided the tails
do not overlap. The spin connection is given by

\beq {(\omega_\mu)}_{ab} \ = \ \sum^N_{i=1} \ \epsilon_{abc} \ \Theta^X_i \
P_i^c
\ \partial_\mu \Theta_i^Y \equiv \sum_{i=1}^N \omega_\mu^{(i)} \label{b14}\eeq
where the $P^a$'s are the ( conserved ) Minkowskian momenta and the $\Theta$-
functions define the particles' trajectories and tails. For instance, for
momenta $P_i$ in the $x$ direction, one can take

$$ \Theta^X_i \ = \ \Theta( V_i T ( x^\mu ) - X ( x^\mu ) ) , $$
\beq \Theta^Y_i \ = \ \Theta ( Y( x^\mu ) - B_i ) \label{b15}\eeq
where $ V_i \ = \ P_i^x / P_i^0 $ ( $P_i^y \ = \ 0$ ) and the $X^a (x)$ are
arbitrary functions of $x^\mu$ , which parametrize the trajectories in the form

\beq  X(\xi_i) \ = \ V_i \ T(\xi_i), \ \ Y(\xi_i ) \ = \ B_i. \label{b16}\eeq

The $\partial_\mu \Theta^Y $ derivative in Eq.
(\ref{b14}) yields a $\delta$-function
singularity on the tails, so that the spin connection is localized, as
anticipated. Since $\omega_i \simeq J \cdot P_i$, it is easy to verify by loop
integration of (\ref{b14}) that the monodromies are given precisely by Eq.
(\ref{b11}).
In particular, if the $X^a$ themselves are chosen as coordinates, one has the
Minkowskian picture of \\ Sec.(\ref{b1}), with the $DJH$ matching
conditions (\ref{b10}).

Corresponding to the spin connection (\ref{b14}), the dreibein solution takes
the form

\beq E^a_\mu \ = \ {( \partial_\mu + \omega_\mu )}^a_b X^b (x) - \sum_i
{( \omega_\mu^{(i)} B_i )}^a \label{b17}\eeq
where the $B_i$'s are the translational parameters occuring in Eq. (\ref{b16}).
Outside the tails, Eq. (\ref{b17}) reduces to
\beq E^a_\mu \ = \ \partial_\mu X^a ( x ) , \ \ \ \ ( {\rm outside \ tails } \
) , \label{b18}\eeq

and the dreibein defines the coordinate transformation from the $x$'s to the
Minkowskian coordinates $X^a$. Our purpose here is to look for single-valued
$x_\mu$'s by restricting the arbitrary functions (\ref{b16}) by a gauge choice.

{\large \bf 2.3 The conformal Coulomb gauge}

According to Eq. (\ref{b18}), the Minkowskian coordinates $X^a$ and the
single-valued
ones $x^\mu \equiv ( t / z /\overline{z} )$, \ \ $z\equiv x+iy$, are related,
outside particle tails, by

\beq dX^a \ = \ E^a_\mu dx^\mu \ = \
A^a dt + B^a dz + {\widetilde{B}}^a d{\bar{z}} ,
\label{b19}\eeq
with the consistency conditions

\beq \partial_{[\nu} E^a_{\mu]} \ = \ 0. \label{b20}\eeq

We shall fix the gauge by a Coulomb condition

\beq \partial_z E^a_{\bar{z}} + \partial_{\bar{z}} E^a_z \ = \ 0.
\label{b21}\eeq
and by a conformal one for the space part of the metric:

\beq g_{zz} \ = \ g_{{\bar z}{\bar z}} \ = \ 0 \label{b22}\eeq

Because of Eqs. (\ref{b20}) and (\ref{b21}) the dreibein components satisfy the
equations

\beq \partial_{\bar{z}} B^a \ = \ \partial_z {\widetilde B}^a \ = \ 0
\label{b23}\eeq
and

\beq \partial_z A^a \ = \ \partial_0 B^a (z,t), \ \ \ \partial_{\bar{z}} A^a
\ = \ \partial_0 {\widetilde B}^a ( {\bar z}, t ) . \label{b24}\eeq

Therefore, $ B^a (z,t) ( {\widetilde B}^a ({\bar z}, t) ) $ are analytic (
anti-analytic ) functions and $ A^a (z,{\bar z},t) $ are harmonic functions
i.e., sums of analytic and antianalytic ones. These analyticity properties
arising from Eq. (\ref{b21}) in two space dimensions, are the counterpart of
instantaneous propagation in a second order formalism \cite{a13},\cite{a14},
and are in fact fundamental to solve the monodromy problem.

Furthermore, because of Eq. (\ref{b22}), $B^a$ and ${\widetilde B}^a$ are null
vectors.
By using straightforward conjugation properties, we can parametrize

\beq  B^a \ = \ N (z,t) \ W^a (z,t) , \ \ \ \  {\widetilde B}^a \ = \ {\bar N}
({\bar z},t) \ {\widetilde W}^a ({\bar z},t) \label{b25}\eeq
where we have defined the null vectors

\beq W^a \ = \ {(f')}^{-1} ( f / 1 / f^2 ) , \ \ \ \ {\widetilde W}^a \ = \
 {({\bar f}')}^{-1} ( {\bar f} / {\bar f}^2 / 1 ) \label{b26}\eeq
in terms of the analytic function $f(z,t)$ that we shall call the mapping
function, of its complex conjugate, and of its derivative $f' (z,t)\equiv df/
dz$.

We can also write

\beq A^a \ = \ ( a / A / {\bar A} ) , \ \ \ a \ = \ {\bar a } , \label{b27}\eeq
where $a (A)$ are real ( complex ) harmonic functions satisfying, by Eq.
(\ref{b24}),
the conditions

$$ \partial_z a \ = \ \partial_t \left( \frac{N}{f'} f \right) , $$
\beq \partial_z A \ = \ \partial_t \left( \frac{N}{f'} \right), \ \ \
\partial_{\bar{z}} A \ = \ \partial_t \left( \frac{N f^2}{f'} \right) .
\label{b28}
\eeq

On the whole, we have now seven real variables $( N, f, a, A )$ in terms of
which we can express, in a straightforward way, the metric tensor in Eq.
(\ref{b13}) as follows:

$$ - 2 g_{z{\bar z}} \equiv e^{2\phi} \ = \ {|\frac{N}{f'}|}^2 {( 1 - {|f|}^2
)}^2 \ = \ {|N|}^2 ( -2 W_a \cdot {\widetilde W}^a ) $$
$$  g_{0z} \equiv \half {\bar \beta} e^{2\phi} \ = \ N W_a \cdot A^a , \
\ \ \ \ g_{0{\bar z}} \ = \ \half \beta e^{2\phi} \ = \ {\bar N}
{\widetilde W}_a  \cdot A^a  $$
\beq g_{00} \ = \ \alpha^2 - {|\beta|}^2 e^{2\phi} , \ \ \ \alpha \ = \ V_a
\cdot A^a \ = \ {\bar \alpha} \label{b29}\eeq
where the $a$ indices are lowered by the minkowskian metric $\eta_{ab}$ with
non-vanishing components $\eta_{00} \ = \ - 2 \eta_{z{\bar z}} \ = \ 1$, and
we have defined the Lorentz vector

\beq V^a \ = \ {(1-{|f|}^2)}^{-1} ( 1+{|f|}^2/ 2{\bar f}/ 2f ) \ = \
\epsilon^a_{bc} W^b {\widetilde W}^c {(W \cdot \widetilde{W})}^{-1}.
\label{b30}
\eeq

Eq. (\ref{b29}) expresses the four real variables of the metric $(\phi, \alpha,
\beta, {\bar \beta})$ in terms of the seven variables of the dreibein. This is
because the metric determines the dreibein only up to local Lorentz
transformations, in this case the three-parameter $O(2,1)$ group. Shortly we
shall take advantage of this fact in order to define a single valued metric.

{\large \bf 2.4 The Mapping Function}

Note now that a non-trivial, i.e, non-Minkowskian metric is obtained because,
due to the matching conditions, the $X^a$ coordinates are multivalued and, in
particular, for $(z - \xi_i) \rightarrow e^{2\pi i} \ (z - \xi_i), \ \
dX^a)_I \rightarrow dX^a)_{II} $, with

\beq dX^a)_{II} \ = \ {(L_i)}^a_b (dX^b)_I , \label{b31}\eeq
according to the matching conditions (\ref{b10}). We should thus require that
the
dreibein components to be multivalued also, and to transform as Lorentz
vectors for loops around the particles' singularities $z \ = \ \xi_i(t)$.

The corresponding metric (\ref{b29}) will be at this point single-valued,
because it
is not affected by a pure Lorentz transformation on the $a$ indices.

This remark suggests a method to solve the monodromy problem. Il $f(z,t)$ has
branch points at $z \ = \ \xi_i(t)$ such that, when $z$ turns around $\xi_i$,
$f$
transforms as a projective representation of the monodromies
(\ref{b11}), then
$W^a$, ${\widetilde W}^a$, $V^a$, will transform as Lorentz vectors by
construction, and so will the dreibein, thus leading to a single-valued metric.

More precisely, by defining a spin-$\half$ representation of the
holonomies (\ref{b11}), i.e.

\beq L_i^{-1} \rightarrow \ell_i^{-1}
\equiv \left( \begin{array}{cc} a_i & b_i \\
{\bar b}_i & {\bar a}_i \end{array} \right) \label{b32}\eeq
with

$$ a_i \ = \ \cos \frac{m_i}{2} + i \gamma_i \sin \frac{m_i}{2} \ , \ \ \ \ \
b_i \ = \ - i \gamma_i {\bar V}_i
\sin \frac{m_i}{2} , $$
\beq \gamma_i \equiv {( 1 - {|V_i|}^2 )}^{-1/2} , \ \ \ \ \
V_i \ = \ ( P^x_i + i P_i^y )/E_i  \ , \label{b33}\eeq
we require that the mapping function $f(z,t)$ transforms as

\beq f(z,t) \rightarrow \frac{a_i f(z,t) + b_i}{{\bar b}_i f(z,t) + {\bar a}_i}
\ , \ \ \ ( i = 1, ... , N ) \label{b34}\eeq

for $ ( z-\xi_i ) \rightarrow e^{2\pi i} ( z-\xi_i )$.

Since the generators of the transformation (\ref{b34}) on analytic (
antianalytic )
functions are

\beq L^a \ = \ \left( f \frac{\partial}{\partial f} / \frac{\partial}{\partial
f} /
f^2 \frac{\partial}{\partial f} \right) , \ \ \ \left[
{\bar L}^a \ = \ \left( {\bar f} \frac{\partial}{\partial {\bar f}} /
{\bar f}^2 \frac{\partial}{\partial {\bar f}}  /
\frac{\partial}{\partial {\bar f}} \right) \right] , \label{b35}\eeq
respectively, it follows that the $W^a$, ${\widetilde W}^a$ in Eq. (\ref{b26})
transform
according to the adjoint ( vector ) representation, and so does $V^a$ in Eq.
(\ref{b30}). It follows that $N(z,t)$ should be single-valued ( i.e., at most
meromorphic, with poles at $z \ = \ \xi_i$) and that $A^a$ will also transform
as a
vector, because of Eqs. (\ref{b24}) and (\ref{b28}).

Our program to solve for the single-valued metric and the corresponding
motion in the conformal Coulomb gauge will thus involve the following steps:

(i) Find the mapping function $f(z,t)$ by solving the monodromy problem in \\
Eqs.(\ref{b32})-(\ref{b34}).

(ii) Find the meromorphic function $N(z,t)$ and the harmonic functions $A^a(z,
{\bar z}, t)$ by integrating the consistency conditions (\ref{b28}) under
proper
asymptotic conditions, to be defined below.

(iii) Find the metric from Eq. (\ref{b29}).

(iv) Find the motion by mapping the Minkowskian trajectories (\ref{b16}), or
more precisely, from the equations

\beq Z(\xi_i,{\bar \xi_i},t) \ = \ B_i + V_i \ T(\xi_i,{\bar \xi_i},t) ,
\label{b36}
\eeq
where the Minkowskian complex velocities $V_i$ are defined in Eq. (\ref{b33})
and $B_i \ = \ B_i^x + i B_i^y$.

\section{ The Two-Body problem}

This is the simplest non-trivial monodromy problem because there are two
non-commuting monodromies, $L_1$ and $L_2$, one for each particle, with

\beq L_i \ = \ e^{\displaystyle{- i J \cdot P_i }} \ , \ \ \ {( i J_a )}_{bc} \
\ = \ \epsilon_{abc}. \label{c1}\eeq

Here the $P_a$'s are the conserved minkowskian momenta which will be assumed
to have a parallel space part, i.e.,

\beq P_i^a \ = \ ( E_i / P_i / {\bar P}_i ) \ = \ m_i \gamma_i \
( 1 / V_i / {\bar V}_i ) \ , \label{c2}\eeq
with $ V_2 \sim V_1 $ ( e.g. $ P_2 = - P_1 $ in the naive c.m. Lorentz frame ).

In order to determine the analyticity properties of the mapping function we
will also assume that initially the tails run outwards and are parallel.
However, it will be clear in a while that the final solutions in single-valued
coordinates will not depend on the fact that the tails may cross, and the
Minkowskian momenta may jump during the motion \cite{a10}. Rather, our
proviso on the tails is to be regarded as an asymptotic initial condition on
the motion, which specifies $P_1^a$, $P_2^a$ and the determination of the
mapping function.

As stated in Sec. 2, we shall first determine the mapping to single-valued
coordinates at a given time $t$, and given particle coordinates $z=\xi_i(t)$,
and we shall find their motion later on. By defining the conformally rescaled
variable

\beq \zeta ( z, t) \ = \ \frac{ z - \xi_1(t)}{\xi_2(t) - \xi_1(t)} \label{c3}
\eeq
the particles' positions are mapped to $\zeta \ = \ 0 $ and $\zeta \ = \ 1$,
around which
the monodromies are given by (\ref{c1}).

However, since in an open universe the composite loop operator around both
particles, for instance $L_{21} \ = \ L_2 L_1$ is non-trivial, then, $\zeta \
= \ \infty$
is also a singularity point of the problem. By explicit computation \cite{a9},
e.g. on the spin $\half$ representation (\ref{b32}) we find

\beq L_2 L_1 \ = \ L_{21} \ = \ e^{\displaystyle{- i P_{21} \cdot J}} , \ \ \ \
\
P^a_{21} \ = \ ( \sqrt{ m^2 + {| p_{21} |}^2} / p_{21} / {\bar p}_{21} ),
\label{c4}\eeq
where

\beq \cos \frac{{\cal M}}{2}  \ = \ \cos \frac{m_1}{2} \ \cos \frac{m_2}{2} -
\frac{ P_1 \cdot P_2}{m_1 m_2} \sin \frac{m_1}{2}  \sin \frac{m_2}{2} ,
\label{c5}\eeq
and

$$ \frac{p_{21}}{\cal M} \sin \frac{{\cal M}}{2} \ = \
\frac{p_1}{m_1} \ \sin \frac{m_1}{2} \ \cos \frac{m_2}{2}
+ \frac{p_2}{m_2} \ \sin \frac{m_2}{2} \ \cos \frac{m_1}{2} + $$
\beq + i (V_2-V_1) \gamma_1 \gamma_2 \sin \frac{m_1}{2} \  \sin \frac{m_2}{2}.
\
\label{c6}\eeq

Thus, note that the monodromy is now dependent on whether we start the
$\zeta \ = \ \infty$ loop anticlockwise on the upper or lower $z$-plane
because $ p_{21} \neq p_{12} $, as it appears from the third term in Eq.
(\ref{c6}).
This is not surprising, due to the non-commutativity. However $L_{12}$ is
related to $L_{21}$ by a similarity transformation which then leaves the
total invariant mass ${\cal M}$ unchanged, i.e., independent of the order of
the
particles.

{\large \bf 3.1 Solution for the mapping function  }

The problem of finding a ( projective ) representation, as in Eq. (\ref{b34})
of
given monodromies, as in Eq. (\ref{b32}), is well known in the mathematical
literature and is referred to as the Riemann-Hilbert problem \cite{a18}. It is
related to the theory of ( second-order ) ordinary differential equations with
Fuchsian singularities.

In fact, if $y_{+} (\zeta)$ and $y_{-} (\zeta)$ are independent solutions of
the differential equation

\beq y'' + q(\zeta) y \ = \ 0 \label{c7}\eeq
with Fuchsian singularities at $\zeta \ = \ \zeta_i$, then it is known
\cite{a17} that
the $\zeta_i$'s are branch points of $y_\alpha (\zeta)$ ($\alpha \ = \ +, -$)
around which they transform linearly according to a subgroup of $SL(2,C)$. If
we are then able to choose $q(\zeta), y_{+}, y_{-},$ such that, for $ (\zeta-
\zeta_i) \rightarrow e^{2\pi i} (\zeta-\zeta_i)$,

\beq \left( \begin{array}{c} y_{+} \\ y_{-} \end{array} \right)
\ \rightarrow \left( \begin{array}{cc} a_i & b_i \\ {\bar b}_i & {\bar a}_i
\end{array} \right) \left( \begin{array}{c} y_{+} \\ y_{-} \end{array}
\right) , \ \ \ \ \ ( i \ = \ 1, ..., N ) \label{c8}\eeq
( where the $a$'s and $b$'s parametrize a spin $\half$ representation of
our $O(2,1)$ monodromies in Eq. (\ref{b33}) ), then the mapping function is
given by

\beq f(z,t) \ \rightarrow \ f( \zeta ) \ = \ \frac{y_{+}(\zeta)}{y_{-}(\zeta)}
\label{c9}\eeq
where we have incorporated the $t$-dependence in the rescaled variable $\zeta
(z,t)$ in Eq. (\ref{c3}).

It is useful to note, for further reference, that in the canonical form
(\ref{c7}),
the "potential" $q(\zeta)$ can be expressed as

\beq 2 q(\zeta ) \ = \ \{ f, \zeta \} \ = \ { \left( \frac{f''}{f'} \right)
}' -
\half { \left( \frac{f''}{f'} \right) }^2 , \label{c10}\eeq
in terms of the Schwarzian derivative $\{ f, \zeta \}$ , which is invariant
under projective transformations
\beq \left\{ \frac{af+b}{cf+d} , \zeta \right\} \ = \ \{ f , \zeta \}.
\label{c11}\eeq

Furthermore, since the Wronskian of two solutions of (\ref{c7}) is constant, it
is easy to realize, by using (\ref{c10}), that a particular basis of solutions
can be expressed in terms of $f(\zeta)$ itself, as follows

\beq Y_{+} \ = \ f( \zeta) {( f'(\zeta) )}^{-1/2} , \ \ \ \
 Y_{-} \ = \ { f'(\zeta) }^{-1/2} , \ \ \ \
( {Y'}_{+} Y_{-} - Y_{+} {Y'}_{-} \ =  \ 1 ). \label{c12}\eeq

For $N \ = \ 2$, we have three Fuchsian singularities, at $\zeta \ = \ 0$,
$\zeta \ = \ 1$,
$\zeta \ = \ \infty$, as remarked before, and the $y$'s are expected to be
expressible in terms of hypergeometric functions. The most general form of
$q(\zeta)$ consistent with the Fuchsian requirements is ( Cfr. Sec. 4 )

\beq q(\zeta) \ = \ \frac{1}{4} \left( \frac{1-\mu_1^2}{\zeta^2} +
\frac{1-\mu_2^2}{{(1-\zeta)}^2} + \frac{1-\mu_1^2-\mu_2^2+\mu_\infty^2}{
\zeta (1-\zeta)} \right) , \label{c13}\eeq
where

\beq \lambda^{\pm}_1 \ = \ \half \ ( 1 \pm \mu_1 ) , \ \ \
\lambda^{\pm}_2 \ = \ \half \ ( 1 \pm \mu_2 ) , \ \ \
\lambda^{\pm}_\infty \ = \ \half \ ( 1 \pm \mu_\infty ) \label{c14}\eeq
represent the pairs of "exponents" at $\zeta \ = \ 0, 1, \infty$ respectively,
which
parametrize the behaviour of the solutions around $\zeta \ = \ \zeta_i$ as
follows
\beq y_\alpha \simeq A^{+}_\alpha (i) {(\zeta-\zeta_i)}^{\lambda_i^{+}} +
A^{-}_\alpha (i) {(\zeta-\zeta_i)}^{\lambda_i^{-}} , \ \ \
( i = 1, 2, \infty, \ \ \ \ \alpha \ = \ \pm ) . \label{c15}\eeq

It is then straightforward to realize that an explicit basis of solutions is
provided by

$$ y_{+} \ = \ k_{+} \zeta^{\half ( 1 + \mu_1 )}
{(1-\zeta)}^{\half ( 1 - \mu_2 )}
{\widetilde F} ( \half (1+\mu_\infty+\mu_1-\mu_2),
\half (1-\mu_\infty+\mu_1-\mu_2);
1+\mu_1;\zeta) , $$
\beq y_{-} \ = \ k_{-} \zeta^{\half ( 1 - \mu_1 )} {(1-\zeta)}^{\half
( 1 - \mu_2 )}
{\widetilde F} ( \half (1+\mu_\infty-\mu_1-\mu_2),
\half (1-\mu_\infty-\mu_1-\mu_2);
1-\mu_1;\zeta) , \label{c16}\eeq
where $y_{-}$ differs from $y_{+}$ for $\mu_1 \rightarrow -\mu_1$ , and we
have defined for convenience a modified hypergeometric function

\beq {\widetilde F} (a,b,c;z) \ \equiv \ \frac{\Gamma(a) \Gamma(b)}{\Gamma(c)}
\ F(a,b,c;z) . \label{c17}\eeq

We also understand,  with the usual determination of F, that the branch cuts of
$y_{\pm}$ run on the intervals $(-\infty,0)$ and $(1,\infty)$, which thus
correspond in the $\zeta$-plane, to the outwards directed initial tails.

We have now to see whether the expressions (\ref{c16}), replaced in Eq.
(\ref{c9}),
match
the monodromy properties of the mapping function. This means that we have to
identify the exponent differences $\mu_1$, $\mu_2$, $\mu_\infty$, and the
coefficient ratio $k\equiv k_{+}/k_{-}$ so as to satisfy the monodromy
conditions (\ref{c8}), or (\ref{b34}).

Since the $F$'s in Eq. (\ref{c16}) are regular at $\zeta \ = \ 0$, it is clear
that
$y_{\pm}$ are relevant solutions in the rest frame of particle \#1, in which
the
Lorentz transformation (\ref{b33}) is diagonal, and reduces to a rotation of
the
mass $m_1$. But $y_{+}/y_{-} \simeq \zeta^{\mu_1}$, so that we can identify

\beq \mu_1 \ = \ \pm \frac{m_1}{2\pi} , \ \ \ \ ( {\rm mod} \ n_1 ) .
\label{c18}\eeq

The monodromy matrix of $y_{+}/y_{-}$ will then be non-diagonal at both
$\zeta \ = \ 1$ and $\zeta \ = \ \infty$ , and determined by the coefficients
in Eq. (\ref{c15}),
which are calculable by analytic continuation \cite{a17}, and given in Table I.
It is clear beforehand that, similarly to Eq. (\ref{c18}), one should have

\beq \mu_2 \ = \ \pm \frac{m_2}{2\pi} \ ( {\rm mod} \ n_2 ) \ , \ \ \ \
\mu_\infty \ = \ \pm \left( -1 + \frac{{\cal M}}{2\pi} \right) , \ \
( {\rm mod} \ 2n-n_1-n_2 ), \label{c19}\eeq
where $\cal M$ is the invariant mass in Eq. (\ref{c5}), characterizing
$\zeta \ = \ \infty$, because the exponent differences $\mu_i$ have an
invariant
meaning which can be established, e.g., by transforming linearly $y_{+}$
and $y_{-}$ to the particle \#2 rest frame, or to an overall c.m. frame,
in which the corresponding monodromy is diagonal.

The determination of $k_{+}/k_{-} \equiv k$ is more subtle, and in a sense
fundamental, because it characterizes the $O(2,1)$ monodromies, compared to
other subgroup of $SL(2,C)$. From the coefficients $A^\beta_\alpha(2)$ in
Table I we find the monodromy matrix

\beq \ell_2^{-1} \ = \ \left( \begin{array}{cc}
\cos\pi\mu_2 + i \sin \pi \mu_2 \frac{\displaystyle{\rho^{+}+\rho^{-}}}{
\displaystyle{\rho^{+}-\rho^{-}}} &
-i \sin \pi \mu_2 \frac{\displaystyle{ 2 \rho^{+}}}{
\displaystyle{\rho^{+}/\rho^{-} - 1}} \\
i \sin \pi \mu_2 \frac{\displaystyle{2 / \rho^{-}}}{
\displaystyle{\rho^{+}/\rho^{-} - 1}} &
\cos\pi\mu_2 - i \sin \pi \mu_2 \frac{\displaystyle{\rho^{+}+\rho^{-}}}{
\displaystyle{\rho^{+}-\rho^{-}}}
\end{array} \right) \ , \ \ \ \rho^\beta \ = \ \frac{A^\beta_{+} (2)}{
A^\beta_{-} (2)} , \label{c20}\eeq
which in general is an $SL(2,C)$ matrix. It matches the $O(2,1)$ form in
eq. (\ref{c8}) provided

\beq  \rho^{+} \ = \
\frac{\gamma_{12} {\bar V}_{21}} {\gamma_{12}-1} \ , \ \ \
1/\rho^{-} \ = \ \frac{\gamma_{12} V_{21}} {\gamma_{12}-1}
\label{c21}\eeq
where the velocities are assumed to be collinear and
$\gamma_{12} \equiv (P_1 P_2)/ m_1 m_2 $ is the $\gamma$-factor of particle
\#2 in the particle \#1 rest frame.

\begin{tabular}{|l|l|l|}  \hline\hline
{\em Table I} &  \multicolumn{2}{c|}{\em behaviour \ at \ $\zeta \ = \  0, 1
$}  \\ \hline & & \\
$\begin{array}{c}  \\ \\ y_{+} \\ \\ y_{-} \end{array} $ &
$\begin{array}{cc} \lambda^{+}_1 & \lambda^{-}_1 \\ & \\
k_{+} \ \frac{\displaystyle{\Gamma(a')\Gamma(b')}}{
\displaystyle{\Gamma(1+\mu_1)}} & 0 \\ & \\
0 & k_{-} \ \frac{\displaystyle{\Gamma(a)\Gamma(b)}}{
\displaystyle{\Gamma(1-\mu_1)}} \end{array} $ &
$\begin{array}{cc} \lambda^{+}_2 & \lambda^{-}_2 \\ & \\
k_{+} \ \Gamma(-\mu_2) & k_{+} \ \frac{\displaystyle{\Gamma(a') \Gamma(b')
\Gamma(\mu_2)}}{\displaystyle{\Gamma(c'-a') \Gamma(c'-b')}}  \\ & \\
k_{-} \ \Gamma(-\mu_2) & k_{-} \ \frac{\displaystyle{\Gamma(a)
\Gamma(b) \Gamma(\mu_2)}}{\displaystyle{\Gamma(c-a) \Gamma(c-b)}}
\end{array} $ \\ & & \\ & & \\ \hline
&  \multicolumn{2}{c|}{\em behaviour \ at \ $\zeta \ = \
\infty$}  \\ \hline  & \\
$\begin{array}{c}  \\ y_{+} \\ \\ y_{-} \end{array} $ & \multicolumn{2}{c|}{
$\begin{array}{cc} \lambda^{+}_\infty & \lambda^{-}_\infty \\ & \\
k_{+} \ e^{\displaystyle{\pm i \pi \mu_1 /2 }} \ \left(
\frac{\displaystyle{\Gamma(b')
\Gamma(\mu_\infty)}}{\displaystyle{\Gamma(1-b)}},
\right. & \left.
\frac{\displaystyle{\Gamma(a')\Gamma(-\mu_\infty)}}{
\displaystyle{\Gamma(1-a)}}
\right) \\ & \\
k_{-} \ e^{\displaystyle{\mp i \pi \mu_1 /2 }} \ \left(
\frac{\displaystyle{\Gamma(b)\Gamma(\mu_\infty)}}{\displaystyle{\Gamma(1-b')}},
\right. & \left. \frac{\displaystyle{ \Gamma(a)
\Gamma(-\mu_\infty)}}{\displaystyle{\Gamma(1-a')}}
\right) \end{array} $ } \\ & \\
\hline  \\ & \\
\multicolumn{3}{|c|}{$ a(\mu_1) \ = \ \half ( 1 + \mu_\infty - \mu_1 - \mu_2 ),
\ \ b(\mu_1) \ = \ \half ( 1 - \mu_\infty -\mu_1 - \mu_2 ) , \ c(\mu_1) \ = \
1 - \mu_1  $ } \\ \multicolumn{3}{|c|}{
$ a' \ = \  a(-\mu_1) , \ b' \ = \ b(-\mu_1), \ c' \ = \ c(-\mu_1) $ }  \\
& \\
\hline \hline
\end{tabular}

By the explicit expressions in Table $I$ we find the conditions

$$ \frac{\rho^{-}}{\rho^{+}} \ = \ \frac{\sin\pi a}{\sin \pi a'} \
\frac{\sin\pi b}{\sin \pi b'} \ = \ \frac{\gamma_{12} - 1}{\gamma_{12} + 1}
\ = \ th^2 \half ( \eta_1 - \eta_2 ) , $$
\beq k \ = \ \rho^{+} \ = \ \frac{\gamma_{12} {\bar V}_{21}}{\gamma_{12} - 1}
\label{c22}\eeq
where $a, b \ ( a', b' )$ are the indices of the $F$'s in Eq. (\ref{c16})
defined in table $I$, and $\eta_i \ = \ th^{-1} |V_i|, \ \ \eta_{12} \ = \
\eta_1-\eta_2$ are the
velocity boosts in a general collinear frame. From the expression
(\ref{c22}) of $k$ , we conclude, as in I, that

\beq f_{(1)} (\zeta) \ = \ \frac{\gamma_{12} {\bar V}_{12}}{\gamma_{12} - 1}
\frac{\zeta^{\mu_1} \ {\widetilde F} (a',b',c';\zeta)}{{\widetilde F}
(a,b,c;\zeta)} \label{c24}\eeq
where the subscript means that it refers to the particle \#1 rest frame.

It is straightforward to check that the condition (\ref{c22}), by Eq.
(\ref{c19}), is
actually equivalent to the definition of the invariant mass in Eq. (\ref{c5}),
thus confirming the expression (\ref{c19}) of $\mu_\infty$. We shall
furthermore
specify Eqs. (\ref{c18}) and (\ref{c19}) by taking the positive signs, or
masses, and by setting $n_i \ = \ n \ = \ 0$, in order to meet the boundary
conditions to be discussed below.

{\large \bf 3.2 Solution for $N$, $A$ and boundary conditions}

The function $N(z,t)$ is single-valued ( meromorphic ) and, according to Eq.
(\ref{b25}), appears as a coefficient of $W^a(f)$ in the expression of $B^a
\ = \ E^a_z$.
Its meaning is better seen in a second-order formalism
\cite{a13} in which its polar
behaviour appears determined by the type of distribution occurring in the
energy-momentum tensor, a single pole corresponding to a $\delta$-function,
a double pole to a $\delta'$, and so on.

In the present approach, in which only global monodromy conditions are set,
the complete determination of $N$ and $A$ requires boundary conditions at the
singularity points, which also help clarifying the determination of the
$\mu_i$ indices in (\ref{c18}) and (\ref{c19}), to be used in the following.

The single particle metric in Eq. (\ref{b8}) shows the following features:

(i) The minkowskian coordinates $Z(\xi_i) \ = \ Z_i$, $T(\xi_i)
\ = \ T_i$ are well
defined in the spinless case, and

(ii) The monodromy behaviour is of type

\beq Z \simeq {(z-\xi_i)}^{1-\displaystyle{\frac{m_i}{2\pi}}} +
Z_i \ \ \ ( |z-\xi_i| << 1 ).
\label{c26}\eeq

Both conditions will turn out to be verified around the singularities if we set
$n_i \ = \ n \ = \ 0$ in Eqs. (\ref{c18}) and (\ref{c19}) and we also take the
ansatz of I, i.e.,

\beq N(z,t) \ = \ \frac{R(\xi (t))}{(z-\xi_1)(z-\xi_2)} \ = \
\frac{R(\xi)}{\xi^2} \frac{1}{\zeta(1-\zeta)} , \label{c27}\eeq
where $\xi \ = \ \xi_2-\xi_1$. In Eq. (\ref{c27}) we have single poles (
corresponding to
$\delta$-function singularities ) and the residues are related so as to avoid
a pole at infinity ( which is unphysical)  and also a zero of the determinant
\beq 2 \sqrt{|g|} \ = \ \alpha \ e^{2\phi} \ = \ {\left| \frac{N}{f'}\right|}
( 1 - {|f|}^2) (V\cdot A). \label{c28}\eeq

We can now discuss the form of the mapping and check the boundary conditions.
By integrating Eq. (\ref{b19}) out of particle \#1, say, we obtain

\beq X^a \ = \ X^a_1 (t) + \int^z_{\xi_1} dz \ N \ W^a (z,t) +
\int^{\bar z}_{{\bar \xi}_1} d{\bar z} \ {\bar N} \
{\widetilde W}^a ({\bar z},t). \label{c29}\eeq

The behaviour of $W^a$ close to the singularity points is better seen by using
the basis \\ (\ref{c12}) and Eq. (\ref{c15}) in the form

\beq W^a ( \zeta ) \ = \ \left( \begin{array}{c} y_{+} y_{-} \\ y_{-}^2 \\
y_{+}^2 \end{array} \right) \simeq \Delta^{(1-\mu)} \ \left(
\begin{array}{c} (A^{-}_{+} + A^{+}_{+} \Delta^\mu)
(A^{-}_{-} + A^{+}_{-} \Delta^\mu) \\ {(A^{-}_{-} + A^{+}_{-} \Delta^\mu)}^2
\\ {(A^{-}_{+} + A^{+}_{+} \Delta^\mu)}^2 \end{array} \right) \label{c30}\eeq
where $\Delta \equiv \zeta-\zeta_i$, we have dropped the $i$ index for
simplicity
, and the coefficients $A^\beta_\alpha(i)$ are given in Table I for $f_{(1)}$.
{}From eqs. (\ref{c30}) and (\ref{c27}) we see that

$$ N W^a \simeq {(\zeta-\zeta_i)}^{-\displaystyle{\frac{m_i}{2\pi}}} \ \
(\zeta \rightarrow \zeta_i , \ \ \ i \ = \ 1,2 ) , $$
\beq \simeq \zeta^{-\displaystyle{\frac{{\cal M}}{2\pi}}} \ , \ \
( \zeta \rightarrow \infty )\label{c31}\eeq

It follows that the endpoint integrals at $z \ = \ \xi_1, \xi_2$ are indeed
well
defined, provided
\beq 0 \le m_{1,2} < 2\pi , \ \ \ \ \ \ \ (8\pi G = 1) \label{c32}\eeq
and that Eq. (\ref{c26}) is verified also.

The point $\zeta \ = \ \infty$ requires further care. Let us first rewrite Eq.
(\ref{c29})
in more detail as follows

\beq X^a \ = \ B^a_1 + V^a_1 T_1(t) +  R (\xi(t)) \ I^a(0,\zeta(z,t))
+ {\bar R} ({\bar \xi}) \ {\widetilde I}^a ( 0 , {\bar \zeta}) ,
\label{c33}\eeq
where

\beq I^a(0,\zeta) \ = \ \int^\zeta_0 \frac{d\zeta}{\zeta(1-\zeta)} W^a (\zeta)
 \ \buildrel {\zeta\rightarrow\infty} \over \simeq \ {(A^{-}_{-})}^2
\zeta^{1-\frac{{\cal M}}{2\pi}}
\left( \begin{array}{c} f(\infty) \\ 1 \\ f^2(\infty) \end{array} \right)
+  A^{-}_{-} A^{+}_{-} \log \zeta \ \left( \begin{array}{c}
\rho^{+}+\rho^{-} \\ 2 \\ 2 \rho^{+} \rho_{-} \end{array} \right)
\label{c34}\eeq
and

\beq \rho^\alpha \equiv \frac{A^\alpha_{+} ( \infty) }{ A^\alpha_{-} (\infty)}
 \ , \ \ \ \ \rho^{-} \ = \ f ( \infty ) . \label{c35}\eeq

We thus see that the behaviour $ \sim \zeta^{1-\frac{{\cal M}}{2\pi}}$ is in
general translated by (\ref{c34}) in the time component $E^a_0 \ = \ A^a$
of the dreibein also, where it would indicate a rotating frame at space
infinity.

More precisely we obtain, by a time derivative of Eq. (\ref{c33})

\beq A^a \ = \ V_1^a {\dot T}_1 + \partial_t [ R(\xi(t)) I^a (0,\zeta(z,t))
+ {\bar R}({\bar\xi} (t)) {\widetilde I}^a (0,{\bar \zeta}(z,t)) ]. \label{c36}
\eeq

Since $\zeta \simeq z/\xi(t)$, the coefficient of the diverging behaviour will
vanish provided

\beq R(\xi) \ = \ \ C \ {\xi(t)}^{1-\displaystyle{\frac{{\cal M}}{2\pi}}} ,
\label{c37}\eeq
thus cancelling the time dependence of the leading term.

We thus see that, as a consequence of the asymptotic condition that $A^a$
be at most logarithmic for $\zeta\rightarrow\infty$, we are able to determine
$R(\xi)$, and thus $A^a$ in Eq. (\ref{c36}), $N$ in Eq. (\ref{c27}) and the
metric in Eq. (\ref{b29}).

The logarithmic behaviour of $A^a$ has a second important consequence. From
Eq. (\ref{c36}) and the asymptotic form (\ref{c34}) we find, after some
algebra, the
asymptotic metric

\beq ds^2 \ \buildrel {|z| >> 1} \over \simeq \  { \left[ d Re \left(
\frac{C}{|\mu_\infty|}
{\xi_{21} (t)}^{1 - \frac{{\cal M}}{2\pi} } \ \log z \right) \right] }^2
- {\rm const. } \ {|z|}^{-\frac{{\cal M}}{\pi}} {|dz|}^2. \label{c38}\eeq

Since $\log z \ = \ \log |z| + i \theta$, it appears from (\ref{c38}) that the
time $T(t,z,{\bar z})$, in a Lorentz c.m. frame, is
asymptotically multivalued, with monodromy related to the particle motion
as follows

\beq T(t, z \ e^{2\pi i}, {\bar z} \ e^{-2\pi i} ) - T(t,z,{\bar z}) \ = \
\frac{2\pi}{
\mu_\infty} Im ( C \xi_{21}^{1-\frac{{\cal M}}{2\pi}} ) , \label{c39}\eeq
where the quantity in the r.h.s. will be related in the next subsection to
the total angular momentum of the system.

{\large \bf 3.3 Solution for the motion}

We have already used the equation of motion for particle \#1 in order to
normalize the inhomogeneous part of $A^a$ in (\ref{c36}) to the time function
$T_1(t)$. The second one will determine the relative motion trajectory
$\xi(t)$, up to some residual gauge freedom.

In fact, from the trajectory equations (\ref{b36}), and the coordinate mapping
(\ref{c33}), we obtain

\beq B_2^a - B^a_1 + T_2 V_2^a - T_1 V_1^a \ = \ C \xi^{1-
\displaystyle{\frac{{\cal M}}{2\pi}}}
I^a(0,1) + {\bar C} {\bar \xi}^{1-
\displaystyle{\frac{{\cal M}}{2\pi}}} {\widetilde I}^a (0,1)
. \label{c40}\eeq

Since $I^a$ and ${\widetilde I}^a$ are calculable constants (see Appendix ),
eq. (\ref{c40}) determines
$\xi(t)$ and the relative time variable $T_1(t) -T_2(t)$, up to an overall time
reparametrization and a scale freedom provided by C.

More in detail, by using a time parametrization in which

\beq T_1(t) \ = \ t - \Delta(t) , \ \ \ T_2(t) \ = \
t + \Delta(t), \label{c41}\eeq
it is straightforward to solve for $\xi$ Eq. (\ref{c40}), in terms of the
relative impact parameter $B$ ( in Minkowskian coordinates ) and of the
integrals
$I^0 , I^z, I^{\bar z}$ and their complex conjugates. After some algebra we
find

$$ C \xi(t)^{1-\frac{{\cal M}}{2\pi}} \ = \ t \ \frac{V_2-V_1}{
I^z + I^{\bar z} - (V_1+V_2) I^0} + i \ \frac{B}{I^z-I^{\bar z}} , $$
\beq \Delta (t) \ = \ t \ \frac{I^0 (V_2-V_1)}{
I^z + I^{\bar z} - (V_1+V_2) I^0} \label{c42}\eeq
where we have assumed that the Minkowskian velocities run along the $x$-axis
( so that $V_1$, $V_2$, and $I^a$ are real ) and the relative impact parameter
is along the $Y$ axis ( so that $iB$ is imaginary ).

Eq. (\ref{c42}) determines completely the form of the mapping in Eq.
(\ref{c33}) in the
time parametrization (\ref{c41}), but does not determine the form of
$\xi_1 (t)$,
which thus remains a residual gauge freedom, in the class of single-valued
solutions. The reason for this freedom is that our gauge choice in Eqs.
(\ref{b20})
and (\ref{b21}) is preserved by holomorphic time dependent conformal
transformations.
However, if we add the boundary conditions (\ref{c26}) at $\zeta \ = \ 0, 1$
and
$\zeta \ = \ \infty$, this freedom reduces to a linear time dependent
transformation
of the form

\beq z \rightarrow a ( z - b(t) ) \label{c43}\eeq
which precisely allows one arbitrary trajectory.

{}From Eq. (\ref{c42}) one can read off the scattering properties. Since
$\xi(t)^{1-\frac{{\cal M}}{2\pi}}$ has the same phase as roughly
$(iB+(V_2-V_1)t)$, the scattering angle is clearly

\beq \theta \ = \ \pm \frac{{\cal M}/2}{1- {\cal M}/2\pi} \ , \ \ \ ( {\rm
sign} \ B \ =  \ \pm ) \label{c44}\eeq
that is, the same as for a test particle \cite{a3} moving in the field of the
total invariant mass, as suggested by 't-Hooft \cite{a5}.

Note that the result (\ref{c44}) is valid for any value of the speed, provided
the coefficient of time keeps its sign, i.e.,

\beq I^z + I^{\bar z} > 2 I^0 > I^0 ( V_1 + V_2 ) , \label{c45}\eeq
which is true because $1 + f^2 > 2f $, $f$ being real for $0<\zeta<1$.
Furthermore it is also independent of the details of the time parametrization
(\ref{c41}), provided monotonicity of time is preserved.

Note also that this result for the scattering angle is much different from the
one found in a covariant gauge of Aichelburg-Sexl type for massless particles
\cite{a10},\cite{a21}.
The instantaneous gauge in the present case forces the particles to interact
at any time, even in the massless limit (Cfr. Sec. 3.5) and shows no sign of a
shock-wave picture.

Finally, let us remark that the result (\ref{c42}) allows the evaluation of the
asymptotic time shift (\ref{c39}). By using the explicit expressions for the
$I^a$'s ( see Appendix ) we find

\beq  I^z - I^{\bar z} \ = \ \frac{\pi \sin \pi \mu_\infty}{\mu_\infty \sin \pi
\mu_1
\sin \pi \mu_2  \gamma_{12} | V_{12} |} \label{c46}\eeq
and therefore, by Eq. (\ref{c39})

\beq \Delta T \ { \buildrel {|z| >> 1} \over \simeq } \
\frac{2\pi}{\mu_\infty} \frac{B}{I^z -
I^{\bar z}} \ = \ 2B \ \frac{\sin \frac{m_1}{2} \sin \frac{m_2}{2} }{\sin
\frac{{\cal M}}{2} } \ \gamma_{12} \ |V_{12}| \ \equiv J \label{c47}\eeq

For small masses, the r.h.s. reduces to $B p_{\rm rel} \ = \ J $, where $p_{\rm
rel}
\simeq \displaystyle{\frac{m_1 m_2}{{\cal M}}} \gamma_{12} |V_{12}|$ is the
relative momentum.
It is thus natural to define the r.h.s. of (\ref{c47}) as the total angular
momentum of
the system for any mass, with a peculiar identification of the "reduced mass"
parameter.

{\large \bf 3.4 Avoiding closed time-like curves}

Having in mind the general two-body dynamics, let us now discuss in more detail
the form of the conformal mapping induced by $f(\zeta)$, and also its possible
breakdown, for particular mass values.

Since the expression (\ref{c24}) of $f_{(1)}$ has no poles in the (upper)
$\zeta$
plane, and has branch cuts at $\zeta \ = \ 0, 1, \infty$ with known behaviour (
Eq.
\ref{c15}), it is straightforward to see that $f_{(1)}(\zeta)$ maps the upper
half $\zeta$-plane into a Schwarz triangle \cite{a20} of type in Fig. $2(a)$.

In drawing Fig. $2(a)$ we have used the boundary values

\beq f_{(1)}(0) \ = \ 0 , \ f_{(1)} (1) \ = \ \rho^{-} (1) \ = \
\frac{\gamma_{12} -1}
{\gamma_{12}+1} , \label{c48}\eeq
and

\beq e^{\mp i\pi \mu_1} f_{(1)}^{\pm} (\infty) \ = \ \rho^{-} (\infty) \ = \
{\left[ \frac{
\sin \displaystyle{\frac{({\cal M} + m_1 - m_2)}{4}}
\sin \displaystyle{\frac{({\cal M} - m_1 - m_2)}{4}} }{
\sin \displaystyle{\frac{({\cal M} - m_1 + m_2)}{4}}
\sin \displaystyle{\frac{({\cal M} + m_1 + m_2)}{4}} } \right]}^{1/2},
( {\rm sign \ Im } \zeta = \pm ), \label{c49}\eeq
which are valid in the mass range

\beq 0 \le m_1, m_2 , \ \ \ \  \ {\cal M} < 2\pi . \label{c50}\eeq

We see that the upper half plane is mapped on a triangle whose edges are
circular arcs, and whose internal angles are $m_1/2$, $m_2/2$ and $\pi -
{\cal M}/2 $, for $m_1+m_2 < {\cal M} < 2\pi$. The lower half plane is
obtained by Schwarz's reflection of this triangle, thus obtaining the
region of Fig. $2(a)$.

It is interesting to note that, in the mass range (\ref{c50}), the whole region
satisfies $|f(z)| < 1$, and the same inequality is satisfied by $f(z)$ on
any other Riemann sheet, because of the elliptic monodromy (\ref{b34}).
Therefore, the determinant of the metric in Eq. (\ref{c28}) is nonvanishing in
the whole $\zeta$-plane.

On the other hand, if $P_1\cdot P_2$ in Eq. (\ref{c5}) exceeds a critical
value,
$\cos({\cal M}/2)$ becomes smaller than $-1$, the mass takes the form
${\cal M} \ = \ 2\pi ( 1 + i \sigma)$, and closed timelike curves appear
\cite{a22}. In
this situation, the behaviour of $f(\zeta)$ for $\zeta\rightarrow\infty$,
provided by Eq. (\ref{c15}), becomes oscillatory because $\mu \ = \ i \sigma$
is pure imaginary and does in fact cross the $|f(z)| \ = \ 1$ value an infinite
number of times, as is apparent from its explicit form ( Fig. $2(b)$ )

\beq e^{\mp i \pi\mu} \ f_{\pm} (\zeta) \
{\buildrel {\zeta\rightarrow\infty} \over \simeq} \
\rho^{-} (\infty) \ \left( \frac{1+{(-\zeta)}^{i\sigma} e^{i\phi_{+}}}
{1+{(-\zeta)}^{i\sigma} e^{i\phi_{-}}} \right) \ , \ \ \
{\cal M} \ = \ 2\pi\left( 1 + i \sigma \right) \label{c51}\eeq
where $|\rho^{-} (\infty)| \ = \  1$, and $\phi_{\pm}$ are proper phases which
can be derived from Table I.

We conclude that the restriction $\cos{\cal M}/2 > -1$ is needed to avoid
both CTC's , and a pathological situation for the gauge choice.

One can
look at the behaviour at $\zeta \ = \ \infty$ also from another point of view.
If
a finite limit $f(\infty)$ exists, in some analyticity sector ( upper and
lower $\zeta$-plane in the present case ), then it must be the same in all
directions of the sector by the Phragm\`en-Lindel\"of theorems \cite{a23}. In
our case
we will have two values, $f_\pm (\infty)$. By applying the monodromies
counterlockwise, we can relate values above and below the cuts as follow

\beq f_{-} (\infty) \ = \ \ell_1 f_{+} (\infty), \ \ \ f_{+} (\infty) \ = \
\ell_2 f_{-} (\infty) . \label{c52}\eeq

We thus find that

\beq f_{+}(\infty) \ = \ \ell_2 \ell_1 f_{+} (\infty), \ \ \  f_{-} (\infty)
\ = \ \ell_1 \ell_2 f_{-} (\infty) , \label{c53}\eeq
that is, $f_{+}(f_{-})$ are fixed points of the composite loop operators
$\ell_{21} ( \ell_{12} )$. By parametrizing it as in (\ref{c4}), we find

\beq V_{21}(\infty) \ f^2_{+}(\infty) - 2 f_{+}(\infty) + {\bar V}_{21}(\infty)
 \ = \ 0 \label{c54}\eeq
where $V_{21} (\infty)\ = \ \frac{p_{21}}{{\cal M}}$ denotes the velocities of
the (upper) c.m. frame in Eq. (\ref{c6}). Therefore,

\beq f_{+}(\infty) \ = \ \frac{1}{V_{21}(\infty)} ( 1 -
\sqrt{ 1 - {| V_{21} ( \infty )|}^2 } ) \label{c55}\eeq
where we have chosen the root such that $|f(\infty)| < 1$. For $V_1 \ = \ 0$,
Eq. (\ref{c55}) reduces to Eq. (\ref{c49}).

However, the square-root in (\ref{c55}) exists properly only if $|V_{12}|, \
|V_{21}|<1$. For too large $P_1 \cdot P_2$ this is no longer the case,
and we end up with $\cos{\cal M}/2 < -1$,
$|f_{\pm}(\infty)| \ = \ 1$, a pathological $\zeta \rightarrow \infty$ limit,
and
closed timelike curves. The nonexistence of a sensible speed for the Lorentz
c.m. frames is yet another hint that such large values of $P_1 \cdot P_2$
have no physical meaning.

{\large \bf 3.5 Massless  and Single Particle Limits}

In our two-body solution we may take the limit of one particle being massless
( $m_2 = 0$, say ) with fixed value of $\gamma_{12} = ( P_1 \cdot P_2 ) / m_1
m_2 $. In this case, the total mass ${\cal M}$ becomes just $m_1$, and we are
describing the single particle limit.

Since $f$ is normalized by Eq. (\ref{c24}) and $N$ by Eq. (\ref{c37}) and
(\ref{c42}) we can check that both $f$ and $N$ vanish, in this limit, so that

\beq \frac{N(z,t)}{f'(z,t)} \ \rightarrow \ {\rm const. } \
{(z - \xi_1 (t) )}^{- \frac{m_1}{2\pi}} \label{c..} \eeq
is a finite quantity. Thus, apart from the already metioned arbitrariness of
$\xi_1 (t)$, we end up with a single particle metric of type

\beq ds^2 \ = \ dt^2 \ - \ {\rm const.} \ {| z - \xi_1 (t) |}^{-\frac{m_1}{
\pi}} {| d ( z - \xi_1 (t) )|}^2 \ , \ \ \eeq
which is just a reparametrization of the static one in Eq. (2.8).

This result, surprising at first sight for a particle that can "move" , is
due to the asymptotic conditions that we have set, which are appropriate for
the configuration space "center of mass" (not to be confused with the Lorentz
ones ) and are in fact inspired to the
single particle metric (\ref{b8}). In other words, if we only have one
particle, it cannot move in an absolute sense, and we end up with a metric
equivalent to the static one.

The massless limit has instead a nontrivial two-body meaning if we let $m_1,
\ m_2 \rightarrow 0$, but also $\gamma_{12} \rightarrow \infty$ so that the
Minkowskian energies $E_1$ and $E_2$ are finite. In such case

\beq \cos\frac{{\cal M}}{2} \ = \ 1 - \frac{{(P_1 \cdot P_2)}}{4} \ = \  1 -
\frac{( E_1 E_2 )}{2} \label{c56}\eeq
provides the invariant mass. In a general frame, with particle boosts
$\eta_1, \eta_2$ one has

\beq f(\zeta) \ = \ \frac{f_{(1)}(\zeta) - th (\eta_1 /2)}{1 - th (\eta_1 /2)
f_{(1)} (\zeta)}. \label{c57}\eeq

We then let $m_i \rightarrow 0$ with $m_i \gamma_i \ = \ E_i$ fixed, so that
$\mu_i \ = \ E_i / 2\pi \gamma_i \rightarrow 0$ and $th \
\eta_1/2 \rightarrow 1 -
m_1 / E_1$ in Eqs. (\ref{c24}) and (\ref{c57}). After some algebra we find, in
a
collinear frame defined by the energies $E_1$ and $E_2$,

\beq f(\zeta) \ = \ \frac{1-f_0 (\zeta)}{1+f_0 (\zeta)} \ , \ \ \
f_0 (\zeta) \ = \ \sqrt{\frac{E_1}{E_2}} \frac{{\widetilde F} ( \half
( 1+\mu_\infty ) , \half
( 1-\mu_\infty ), 1, 1-\zeta )}{{\widetilde F} (
\half ( 1+\mu_\infty ) ,
\half ( 1-\mu_\infty ), 1, \zeta )} \ , \label{c58}\eeq
where $\mu_\infty \ = \ {\cal M}/{2\pi} -1$ as usual.

Note that both hypergeometric functions have $c-a-b \ = \ 0$, so that they
provide
logarithmic singularities at $\zeta \ = \ 0$ and $\zeta \ = \ 1$, as expected.
The
behaviour at $\zeta \ = \ \infty$ is normal, and we have to require $E_1 E_2 <
4$
in order to avoid CTC's and a pathological metric.

We also obtain, from Eq. (3.45), the limiting form of the asymptotic time
shift in the massless case

\beq \Delta T \ = \ J \ = \ 2 B tg \frac{{\cal M}}{4} \ = \ B {\cal E}
\label{c59}\eeq
where we have defined the "effective energy" $ {\cal E} \ = \ 2 tg (
\frac{{\cal M
}}{4} )$. It is amusing to note that in the Aichelburg-Sexl gauge of Ref.
\cite{a10} the massless scattering angle and energies are

\beq \theta_{A.S.} \ = \ \frac{{\cal M}}{2} \ , \ \ \ E_{A.S.} \ = \ {\cal E}
\ = \ 2 tg \frac{{\cal M}}{4} \label{c60}\eeq
and that, therefore, the bound ${\cal M} < 2\pi $ is built in, in the physical
energy range.

In the limiting case ${\cal M} \ = \ 2\pi$ (which requires care in most
formulas ),
the mapping function $f_0 (\zeta)$ reduces to the well known function

\beq f_0 \ = \ \sqrt{\frac{E_1}{E_2}} \frac{{\widetilde F} (
1/2,1/2,1;1-\zeta)}{
{\widetilde F} ( 1/2, 1/2; 1; \zeta)} \ , \label{c61}\eeq
which is the inverse of the automorphic function $\zeta \ = \ \kappa^2 (\tau)$,
occuring in the theory of elliptic integrals [20].

\section{Main Features of the N-Body Problem}

Our method of solution for the metric works for $N\ge 3$ as well, once the
mapping function is found. However, unlike the $N \ = \ 2$ case, we have found
no
explicit general form of the mapping for $N \ge 3$, except to first order in
the relative speed , i.e. in the "quasi static case" , that will be discussed
in a moment. The main difficulty that we find is the fact that the Fuchsian
problem for the mapping shows additional "apparent singularities"
besides the $N+1$ expected ones. For instance, the three-body problem requires
the solution of a Fuchsian equation with five singularities, which is not
explicitly known.

In this section, we set up the general problem and we solve the quasi-static
case. In the next, we discuss some explicit examples that we are able to treat
in full detail.

{\large \bf 4.1 The Fuchsian Problem}

The method for finding the mapping function works exactly as in the two-body
case, and is based on the fuchsian differential equation (\ref{c7}), i.e.

\beq y'' + q(\zeta) y \ = \ 0 . \label{d1}\eeq

However, the potential $q(\zeta)$ is now dependent on some set of singularities
$\zeta \ = \ \zeta_i$ (i \ = \ 1,...,n), yet to be found. Because of the
Fuchsian
constraints, we can parametrize $q(\zeta)$ in terms of double and single
poles at the singularities, i.e.

\beq 2q(\zeta) \ = \ \sum^n_{i=1} \left( \frac{1-\mu^2_i}{2{(\zeta-\zeta_i)}^2}
+ \frac{\beta_i}{\zeta-\zeta_i} \right) \ , \label{d2}\eeq
where

\beq \lambda^{\pm}_i \ = \ \half ( 1 \pm \mu_i ) \label{d3}\eeq
are the exponents at the singularities, and the residues $\beta_i$ are
accessory
parameters which could accomodate our nonabelian monodromies.
The condition that $\zeta \ = \ \infty$ be a Fuchsian singularity yields two
constraints on the $\beta$'s , i.e. ,

\beq \sum^n_{i=1} \beta_i \ = \ 0 \ , \ \ 1-\mu^2_\infty \ = \ \sum^n_{i=1}
( 1 - \mu^2_i + 2 \beta_i \zeta_i ) \ , \label{d4}\eeq
where $\mu_\infty$ is the corresponding difference of exponents.

In the two body case, the particles can be set at $\zeta \ = \ 0$ and $\zeta \
= \ 1 $ by
the conformal transformation (\ref{c3}) and $\beta_1 , \ \beta_2$ are
completely
determined by (\ref{d4}) to be

\beq \beta_1 \ = \ - \beta_2 \ = \ \half ( 1 + \mu^2_\infty - \mu^2_1 -
\mu^2_2 ) , \label{d5}\eeq
so that the potential (\ref{c13}) emerges. No additional singularity is needed,
because the number of invariants required by the momenta is precisely three
\footnote{There is a hidden parameter, the ratio of coefficients of two
independent solutions, which however is needed to match the $O(2,1)$ nature
of the monodromies (Cfr. Sec. 3.1).}.

In the general case, we have $N$ three-momenta, and $3N-3$ invariants.
Assuming $N$ particle singularities and one at $\zeta \ = \ \infty$ we only
have
$2N-1$ free parameters, $N-2$ coming from the $\beta$'s. Thus, we need
additional singularities which yield trivial monodromy properties for the
mapping function. These are the "apparent singularities" \cite{a18}.

Since an apparent singularity $\zeta_j \ ( j \ge N+1 )$ has trivial monodromy,
the difference of exponents $\mu_j$ should be an integer, the simplest case
being $\mu_j \ = \ 2$ ( $\mu_j \ = \ 1$ means no singularity ). A $\mu_j  \ = \
2$
singularity, corresponding to exponents $-1/2$ and $3/2$ (Eq. (\ref{d3})),
yields a simple zero of $f'$, because one of the solutions is $ \simeq
{(f')}^{-1/2} $ ( Eq. (\ref{c12})).

Setting $\mu_j \ = \ 2$, say, is however not enough to insure the absence of
$\log ( \zeta-\zeta_j )$ terms in the solution, which yield nontrivial
monodromy. Around $\zeta \ = \ \zeta_j$ we can write

\beq 2q(\zeta) \ = \ \frac{-3/2}{{(\zeta-\zeta_i)}^2} + \frac{\beta_j}{\zeta-
\zeta_j} + 2 {\bar q}_j ( \zeta ) \ , \ \ ( j \ = \ N+1, ..., n )
\label{d6}\eeq
where ${\bar q}_j$ is regular at $\zeta_j$. Then a simple analysis shows
that the "non-logarithmic condition" is

\beq -\half \beta^2_j \ = \ 2 {\bar q}_j ( \zeta_j) \ = \ \sum_{i\neq j}^n
\left[ \frac{1-\mu^2_i}{2{(\zeta_j-\zeta_i)}^2} +
\frac{\beta_i}{\zeta_j-\zeta_i} \right] \ , \ \ \ ( \mu_j \ = \ 2 ).
\label{d7}\eeq

By assuming now that all apparent singularities have $\mu_j \ = \ 2$, i.e. ,
are
simple zeros of $f'$ ( multiple zeros being a limiting case of this one )
and satisfy the nonlogarithmic condition (\ref{d7}), it is easy to realize that
we need $N-2$ of them to solve our problem.

In fact, having fixed the $N+1$ exponents

\beq \mu_i \ = \ \frac{m_i}{2\pi} , \ \ \ i \ = \ 1, ...., N \ ; \
\mu_\infty \ = \ \frac{{\cal M}}{2\pi} - 1 \ , \label{d8}\eeq
we need $2(N-2)$ parameters to accomodate $N-2$ complex relative velocities.
On the other hand, we have $(N-2)$ normal residues $\beta_i \ \ \
( i \ = \ 3,...,N) $ , and $(n-N-1)$ residues $\beta_j$ and positions
$\zeta_j \ (j \ = \ N+1,..., n)$ of the apparent singularities, of which only
one
per singularity is to be counted, because of the nonlogarithmic conditions.
This determines

\beq n \ = \ 2 N - 1 \label{d9}\eeq
and the number of apparent singularities to be $(N-2)$.

Note that we have not counted as parameters the $N-2$ particle positions
$\zeta_i (t) , ( i \ = \ 3,..N ) , $ because the latter should be determined
dynamically from the equations of the motion of the problem. In order to
distinguish them from the apparent singularities, we shall sometimes use for
the latter the notation $\eta_k \ = \ \zeta_{k+N}, \ \ \gamma_k \ = \
\beta_{k+N}
, $ with $k \ = \ 1, ..., N-2$. Then, the above counting of conditions and
parameters means that, given the set of particle singularities

\beq 0, \ 1, \ \zeta_3, ..., \zeta_N, \ \infty , \label{d10}\eeq
and corresponding exponents, the residues $\beta_{2+i}$ and $\gamma_k \ ( i, k
\ = \ 1, ...., N-2 ) $ should be determined from the relative velocities in the
monodromy matrices, while the apparent singularities $\eta_k ( \beta_i,
\gamma_i;\zeta_i )$ would then be dependent variables because of the
non-logarithmic conditions.

The really awkward part of this program is the determination of the monodromy
matrices by an analytic continuation of the solutions, which however is
not yet explicitly available. General results are instead available in the
mathematical literature  on the functional dependence $\eta_k(\zeta_i)$, at
constant monodromy matrices $L_i(P_i)$.

In fact, if we decide to solve the $(N-2)$ non-logarithmic conditions
(\ref{d7})
for the $\beta_{2+i} ( \gamma_k, \eta_k)$ ( because they are linear in such
quantities ) , we obtain \cite{a24},\cite{a18}
 the so called "Garnier systems", hamiltonian
systems in which $\{ \eta_k, \gamma_k \}$ is a set of conjugate variables, and
$\{ \zeta_{i+2}, \beta_{i+2} \}$ is a corresponding set of time-hamiltonian
pairs,
with equations:

\beq \frac{\partial \eta_k}{\partial \zeta_{i+2}} \ = \
\frac{\partial \beta_{i+2}
( \gamma , \eta )}{\partial \gamma_k} \ , \ \
\frac{\partial\gamma_k}{\partial\zeta_{i+2}} \ = \ - \frac{\partial\beta_{i+2}
(\gamma,\eta)}{\partial\eta_k} , \ \ (i, k \ = \ 1, ..., N-2). \label{d11}\eeq

This (quite nonlinear) set of equations with $N-2$ "times" $\zeta_3,...,
\zeta_N$ constrains the dependence $\eta_k(\zeta_{i+2})$ so as to insure that
the Minkowskian momenta are constants of motion. For $N \ = \ 3$, the system
(\ref{d11}) reduces to the VI-th Painlev\`e equation \cite{a25} for
$\eta(\zeta_3)$.

Thus, we conclude that the monodromy problem for $N \ge 3$ is in principle
solvable, but the general solution is not yet explicit.

{\large \bf 4.2 The Quasi Static Case}

The $N$-body problem can be explicitly solved to first order in the
velocities of the particles. In this case, the singularity exponents turn
out to be the static ones, but the particles can move, and the solution
shows rather nontrivial features.

The basic observation is that the mapping function is itself of first order
in the velocities, as it can be seen from Eq. (\ref{c24}), by taking care of
the
fact that $a\rightarrow 0$ in this limit. Therefore, the projective
transformation (\ref{b34}) linearizes in the form

\beq f(\zeta) \rightarrow e^{2i \pi \mu_i} f + \frac{{\bar V}_i}{2} (
e^{2i \pi \mu_i} - 1). \label{d12}\eeq
for $(\zeta-\zeta_i) \rightarrow e^{2i\pi} (\zeta-\zeta_i).$

Although (\ref{d12}) is still non commutative, it becomes trivially commutative
for the derivative, i.e.,

\beq f' (\zeta) \rightarrow e^{2i\pi \mu_i} f' (\zeta) . \label{d13}\eeq

We shall thus solve (\ref{d13}) first, in the form

\beq f' (\zeta) \ = \ K \prod^{2N-1}_{i=1} {(\zeta-\zeta_i)}^{\mu_i-1} \ = \
K \prod^N_{i=1} {(\zeta-\zeta_i)}^{\mu_i-1} \cdot \prod^{N-2}_{k=1}
(\zeta-\eta_k) , \label{d14}\eeq
where $\zeta_1 \ = \ 0, \ \ \zeta_2 \ = \ 1$,

\beq \mu_i \ = \ \frac{m_i}{2\pi}, \ \  (i \ = \ 1,...,N), \ \
\mu_i \ = \ 2, \ \ ( i \ = \ N+1, ...,2N-1 ), \label{d15}\eeq
and we have explicitly shown the zeros at $\zeta \ = \ \eta_k \equiv
\zeta_{k+N}$.

Then, we obtain the mapping function $f_{(1)} (\zeta)$ in the particle \#1
rest frame in the form

\beq f_{(1)} (\zeta) \ = \ K \int^\zeta_0 d\zeta \prod_{i=1}^{2N-1} {(\zeta-
\zeta_i)}^{\mu_i-1} , \label{d16}\eeq
where the cuts at the branch points are assumed to run outwards and to not
overlap, for a given cyclic initial ordering of the particles (Fig. 3).

The function (\ref{d6}) changes by just the phase $exp(im_1)$ around particle
\#1,
but has nontrivial monodromies around the remaining ones. In order to find
them, we use additivity of the integrals in their analyticity domain, to
write

\beq f_{(1)} (\zeta) \ = \ f_{(1)}(\zeta_i) + f_{(i)}(\zeta) . \label{d17}\eeq

Since $f_{(i)}(\zeta)$ changes by $exp(i \ m_i)$ around $\zeta_i$, we find
the monodromy at $\zeta_i$ to be

\beq f_{(1)}(\zeta) \rightarrow e^{i m_i} f_{(1)}(\zeta) + f_{(1)}
(\zeta_i) ( 1 - e^{i m_i} ) . \label{d18}\eeq

By matching (\ref{d18}) to (\ref{d12}) we find the conditions

\beq f_{(1)} (\zeta_i) \ = \ K \int^{\zeta_i}_0 d\zeta \prod_{i=1}^N
{(z-\xi_i)}^{\mu_i-1} \prod_{k=1}^{N-2} (z-\eta_k) \ = \ \frac{{\bar V}_1 -
{\bar V}_i}{2} , \label{d19}\eeq
where we have restored $V_1$, in a general frame. The $(N-1)$ equations
(\ref{d19}) determine $K(\zeta_i)$ and $\eta_k(\zeta_i)$ in terms of the
$V_i$'s and of the particle positions $\zeta_3, ...,\zeta_N$. Note that we
could not have matched the monodromies without the zeros $\eta_k$, which
seemed perhaps useless in Eq. (\ref{d14}).

Of course, since we have the solutions (\ref{d14}) and (\ref{d16}), all
parameters of
the potential $q(\zeta)$ are also determined. By using the relation (\ref{c10})
with the Schwarzian derivative, we find

\beq 2q(\zeta) \ = \ {\left( \frac{f''}{f'} \right) }' - \frac{1}{2}
{\left( \frac{f''}{f'} \right) }^2 \ = \ \sum^{2N-1}_{i=1}
\left( \frac{
(1-\mu_i^2)}{2{(\zeta-\zeta_i)}^2} + \frac{\beta_i}{\zeta-\zeta_i} \right)
, \label{d20}\eeq
where

\beq \beta_i \ = \ - (1-\mu_i) \sum^{2N-1}_{j\neq i} \frac{1-\mu_j}{\zeta_{ij}}
\ , \ \ \ ( \zeta_{ij} \equiv \zeta_i - \zeta_j ) . \label{d21}\eeq

The accessory parameters just found satisfy $\sum_i \beta_i \ = \ 0$
identically,
and define $\mu_\infty$ by the equation (\ref{d4}), yielding after a simple
algebra

\beq \mu_\infty \ = \ - \sum^{2N-1}_{i=1} (1-\mu_i) + 1 \ = \ \sum^N_{i=1}
\mu_i
- 1 , \label{d22}\eeq
thus confirming the expectation that the invariant mass is just static, to
first order in $V_i$'s.

Finally, one can also check with some algebra that the expressions (\ref{d21})
satisfy the $(N-2)$ nonlogarithmic conditions (\ref{d7}).
Eqs. (\ref{d16}), (\ref{d19}) and (\ref{d21}) represent the complete solution
of the
mapping problem in the quasi-static case. Their explicit expressions for
$N=3$ are given in Sec. (5.1).

The qualitative form of the mapping induced by $f(\zeta)$ is pictured in
Fig. 4  for $N=3$, where the values $f_{i i+1}(\infty)$ are discussed in the
next subsection.

{\large \bf 4.3 Metric and motion}

Given the mapping function, we have still to find $N$ and $A$. Since the
zeros of $f'$ would imply poles in $NW^a$ , we require the function $N$ to
cancel them. By assuming $N$ to have zeros at $\zeta \ =
\ \eta_k$ , simple poles
at $\zeta \ = \ \zeta_i$, ( as for $N \ = \ 2$ ) and no pole at $\zeta \ = \
\infty$, we obtain

\beq N(z,t) \ = \ \frac{R(\xi_{21},\zeta_i)}{\xi^2_{21}} \
\frac{\prod_{k=1}^{N-2} (\zeta-\eta_k)}{\prod^N_{i=1} (\zeta-\zeta_i)} , \ \ \
\ (\xi_{21} \ = \ \xi_2-\xi_1) , \label{d23}\eeq
where $\zeta \ = \ (z-\xi_1)/\xi_{21}$ is again the rescaled variable
(\ref{c3}), and $\zeta_i \ = \ \xi_{i1}/\xi_{21}$. The form \\
(\ref{d23}) satisfies the boundary conditions (i) and (ii) of Sec. (3.2).

We can then write the detailed form of the mapping as in Eq. (\ref{c33}),
namely

\beq X^a \ = \ B_1^a + V_1^a T_1(t) + R(\xi_{12},\zeta_i) I^a(0,\zeta(z,t)) +
{\bar R} {\widetilde I}^a (0, {\bar \zeta}) , \label{d24}\eeq
where now

\beq I^a(0,\zeta) \ = \ \int^\zeta_0 \ d \zeta \frac{\prod(\zeta-\eta_k)}{
\prod(\zeta-\zeta_i)} W^a(\zeta). \label{d25}\eeq

By inserting in (\ref{d25}) the asymptotic behaviour of $f' ( \zeta)$ that we
write in the form

\beq f' (\zeta) \ { \buildrel {\zeta\rightarrow\infty} \over
\longrightarrow } \ K(\zeta_i) \
\zeta^{\frac{{\cal M}}{2\pi}-2} ( 1 - \frac{{\cal M}}{2\pi} )
, \ \ \ \ \label{d26}\eeq

we find
\beq I^a(0,\zeta) \ { \buildrel {\zeta\rightarrow\infty} \over \sim } \
{\left(\frac{z}{\xi_{21}} \right)}^{1 - \frac{{\cal M}}{2\pi}}
\frac{1}{K(\zeta_i)} \left( \begin{array}{c} f(\infty) \\ 1 \\ f^2(\infty)
\end{array} \right) , \label{d27}\eeq
where $f(\infty)$ is the asymptotic value of $f(\zeta)$ in one of its
analyticity sectors (Fig. 3), to be discussed shortly.

{}From Eqs. (\ref{d24}) and (\ref{d27}) it follows that $A^a$ is badly behaved
at
infinity, unless we set, similarly to the two-body case

\beq R(\xi_{12}, \zeta_i) \ = \ \ C \ K(\zeta_i)
{(\xi_{21})}^{1-\frac{{\cal M}}{2\pi}}. \label{d28}\eeq

We see that the asymptotic condition $A^a \sim \log z$ determines this time
the dependence of $R$ on both $\xi_{12}$ and $\zeta_i$, in terms of
$K(\zeta_i)$
in Eq. (\ref{d26}).

As a consequence, the metric is fully determined and so are the equations of
motion

\beq B_i^a - B_1^a + V_i^a T_i - V_1^a T_1 \ = \ C \ \xi_{21}^{1-\frac{{\cal
M}}
{2\pi}} K\left(\frac{\xi_{i1}}{\xi_{21}}\right)
I^a\left(0, \frac{\xi_{i1}}{\xi_{21}}\right) +
{\bar C} {\bar \xi}_{21}^{1-\frac{{\cal M}}{2\pi}} {\bar K}
{\widetilde I}^a \ \ \ \ (i \ = \ 2,...,N). \label{d29}\eeq

These $(N-1)$ equations determine the relative times $T_i-T_1$ and the
relative motion trajectories $\xi_{i1}(t)$. Of particular interest is the
motion in the dimensionless parameters $\xi_{i1}/{\xi_{21}}$ discussed in
Sec. (5.1) for $N \ = \ 3$.

So far, our discussion on $N$ and $A$ has been general, and applies to the
$N$-body case for any speed. In the quasi-static case, the invariant mass
$\cal M $ takes the static value of \\ Eq.(\ref{d22}) and $K(\zeta_i)$ is
determined by the eqs. (\ref{d19}).

Furthermore, the value of $f(\infty)$ in Eq. (\ref{d27}) can be determined as
follows. As in the
two-body case ( Sec. 2.4) , we expect $N$ different values of $f(\infty)$ in
the various analyticity sectors $ (N1), \ (12), ...., \ (N-1,N)$ of Fig. (3).
They correspond to the fixed points of the various monodromies at infinity
that we can have, for instance, to

\beq \ell_N ... \ell_1 \ f^{(N_1)} (\infty) \ = \ f^{(N1)} (\infty) ,
\label{d30}
\eeq
and , for the others, to

\beq f^{(12)} (\infty) \ = \ \ell_1 \ f^{(N1)} (\infty), .... \label{d31}\eeq
and so on.

Referring to Eq. (\ref{d30}) for definiteness, and combining the monodromies
(\ref{d12}),
we easily obtain

\beq f^{(N1)} (\infty) \ = \ - \frac{{\bar V} (N1)}{2}, \label{d32}\eeq
where

\beq V(N1) \ = \ \frac{ \prod_2^N e^{im_i} ( e^{im_1} -1 ) V_1 +
\prod_3^N e^{im_i} ( e^{im_2} -1 ) V_2 + ... ( e^{im_N} -1 ) V_N }{
\prod_1^N e^{im_i} - 1} \label{d33}\eeq
is the velocity of the corresponding "center-of-mass" Lorentz frame. We have
thus $N$ c.m. systems and $N$ values at $\zeta \ = \ \infty$, for a given
initial
cyclic ordering. This fact was used in drawing the qualitative mapping of
Fig. (4), and is a consequence of the noncommutativity of the monodromies,
even in this simplified limit.

Since the velocities are assumed to be small, we have in principle no
problems with the requirement $|f(\infty)|<1$ , needed to have a
nonsingular mapping. However, Eq. (\ref{d33}) shows that something wrong
happens
if $\sum_i m_i \ = \ 2\pi$, because the total velocity shows a pole. We thus
expect that a condition similar to ${\cal M} < 2\pi$ should be imposed in
general.

Note, however, that for $N\ge 3$ and finite velocities even the invariant
mass will be dependent on the inequivalent cyclic orderings of the particles
which insure initially nonoverlapping tails. This is yet another feature of
the general $N$-body problem which requires further study.

\section{Some explicit solutions}

{\large \bf 5.1 The Quasi-static three-body case}

The detailed study of the $N \ = \ 3$ motion shows, even to first order in the
velocities, some features of general interest, for instance related to the
decoupling limit of the two-body subsystems from the three body one, which
are worth looking at in detail.

In the $N \ = \ 3$ case there is only one
nontrivial position $\zeta_3 \ = \ \xi_{31} / \xi_{21}$, for particle \# 3, and
one apparent singularity $\eta(\zeta_{3})$, that we can determine explicitly.
The total number of singularities is thus five. As a consequence, the
system of two equations in (\ref{d19}) is linear in $\eta$ and has the
following
solution

\beq \eta(\zeta_{3})-\zeta_{3} \ = \ \frac{{\bar V}_{31} I_{12} (\mu_3 + 1,
\zeta_3) -
{\bar V}_{21} I_{13} ( \mu_3 + 1, \zeta_3 )}{{\bar V}_{31} I_{12} (\mu_3,
\zeta_3)
- {\bar V}_{21} I_{13} ( \mu_3, \zeta_3 )} \ = \ \frac{\sum {\bar V}_i I_{jk}
( \mu_3+1, \zeta_3 )}{\sum {\bar V}_i I_{jk} ( \mu_3, \zeta_3 )} \label{e1}\eeq
where we have used the notation

\beq I_{ij} ( \mu_3, \zeta_3) \ = \ \int^{\zeta_j}_{\zeta_i} dz \ z^{\mu_1 - 1}
\
{(z-1)}^{\mu_2-1} {(z-\zeta_3)}^{\mu_3-1} \ , \ \ \ V_{ij} \equiv V_i - V_j .
\label{e2}\eeq

The corresponding solution for $K$ is

\beq K(\zeta_{3}) \ = \ \frac{\mu_3}{2} \frac{ \sum_{\rm cyclic} {\bar V}_i
I_{jk} ( \mu_3, \zeta_3 )}{ W [ I_{12} ( \mu_3 + 1, \zeta_3 ), I_{13}
( \mu_3 + 1, \zeta_3 ) ]} \ , \label{e3}\eeq
where $W( y_1, y_2 ) \equiv  {y'}_1 y_2 - y_1 {y'}_2$ is a Wronskian.

The integrals $I_{12} ( \mu_3 + 1, \zeta )$ and $I_{13} ( \mu_3 + 1, \zeta )$
are expressible in terms of hypergeometric functions with singularities at
$\zeta \ = \ 0, 1 , \infty$ ( the remaining particle singularities ) and
difference
of exponents $\mu_1+\mu_2 , \ \mu_2 + \mu_3, \ \mu_1 + \mu_2 - 1$ respectively.
Their explicit form, for the branch-cut structure of Fig. (3) is

$$ I_{12} ( \mu_3 + 1, \zeta ) \ = \ e^{ i \pi ( \mu_2 -1)} B(\mu_2, \mu_1 +
\mu_3 ) F ( \mu_3, 1 - \mu_1 - \mu_2 - \mu_3 ; \zeta ) $$
$$ - e^{i \pi ( \mu_2 - \mu_1 - \mu_3 )} B(\mu_1 , - \mu_1 - \mu_3 )
\zeta^{\mu_1 + \mu_3} F( \mu_1, 1 - \mu_2 , 1 + \mu_1 + \mu_3 ; \zeta ) ,  $$
\beq I_{13} ( \mu_3 + 1, \zeta ) \ = \ - e^{i \pi ( \mu_2 - \mu_3 )}
B( \mu_1 , 1 + \mu_3 ) \zeta^{\mu_1 + \mu_3} F( \mu_1, 1 - \mu_2, 1+ \mu_1 +
\mu_3 ; \zeta ) , \label{e4}\eeq
and we also have the Wronskian

\beq W \ = \ e^{i \pi ( 2 \mu_2 - \mu_3 )} \frac{ \Gamma ( \mu_1 ) \Gamma (
\mu_2 )
\Gamma ( 1 + \mu_3 ) }{ \Gamma ( \mu_1 + \mu_2 + \mu_3 ) } \zeta^{\mu_1 + \mu_3
- 1} \ { ( 1 - \zeta )}^{\mu_2 + \mu_3 - 1} , \label{e5}\eeq
in terms of which $\eta ( \zeta_3 )$ and $K ( \zeta_3)$ are explicitly found.

An interesting point to notice in Eq. (\ref{e1}) is the limiting behaviour of
$\eta (\zeta_3)$ for $\zeta_3$ close to the singularity points $ 0, 1 ,
\infty$.
It is easy to check that they approach the same limit. For instance if

\beq \zeta_3 \ = \ \frac{\xi_{31}}{\xi_{21}} \rightarrow 0 \ \ \ \
\eta \ = \ \frac{\xi_{41}}{\xi_{21}} \simeq \ \ {\left(
\frac{\xi_{31}}{\xi_{21}}
\right)}^{1 -\mu_1 - \mu_3} \rightarrow 0 . \label{e6}\eeq

This means that if $\xi_3$ becomes degenerate with $\xi_1$, the location of the
zero does too. This is needed in order to have a correct two-body limit of the
mapping function. In fact, in the same limit we have

\beq f' ( \zeta ) \rightarrow \ K( \zeta_3 ) \ \zeta^{\mu_1 + \mu_3 -1}
{(\zeta - 1)}^{\mu_2 - 1} \ , \label{e7}\eeq
as expected from a system of masses $m_1 + m_3$ and $m_2$. Correspondingly,
$f$ and $N$ have the behaviour typical of three singularities, instead of
five: one pole and one zero have disappeared. This behaviour is expected to
hold
for general velocity configurations.

One can check that the expression (\ref{e1}) of $\eta(\zeta_3)$, that we have
found by
solving explicitly for the monodromies ( Eq. \ref{d19} ), is also a solution
of the VI-th Painlev\`e equation for the value $\mu_{\infty} \ = \
\sum^3_{i=1} \mu_i - 1 $ of the exponent at infinity. This nonlinear
second-order
equation comes from the fact that the monodromies do not change when $\zeta_3$
varies ( "isomonodromic problem" ) \cite{a25} and is therefore a consistency
check of the present approach.

Let us now look in more detail to the three-body relative motion. According
to \\
Eq.(\ref{d29}) we have now two equations in which, to first nontrivial order
in $v$, we have, by (\ref{d25}),

\beq R \ I^a ( 0 , \zeta_i ) \ = \ \ C {( \xi_{21} )}^{1- \frac{{\cal M}}{2\pi}
}
\ \int^{\zeta_i}_0 \ d\zeta \zeta^{-\mu_1} {( \zeta - 1)}^{-\mu_2}
{(\zeta - \zeta_3 )}^{-\mu_3} \left( \begin{array}{c} O(v) \\ 1 \\ O(v^2)
\end{array} \right) . \label{e8}\eeq

{}From Eq. (\ref{d29}) we then obtain $T \ = \ t + O(v^2)$, and

$$ Z^0_{21} + V_{21} t \ = \ C \xi_{21} (t)^{1- \sum \mu_i} \int^1_0 \
d\zeta \zeta^{-\mu_1} {( \zeta - 1 )}^{-\mu_2} {(\zeta - \zeta_3(t))}^{-\mu_3}
$$

\beq Z^0_{31} + V_{31} t \ = \ C \xi_{21} (t)^{1- \sum \mu_i} \int^{\zeta_3}_0
d\zeta \zeta^{-\mu_1} {( \zeta - 1 )}^{-\mu_2} {(\zeta - \zeta_3(t))}^{-\mu_3}
\label{e9}\eeq

It follows, remarkably, that the $\zeta_3$ motion decouples from that of
$\xi_{21}$, in the form

\beq \frac{Z^0_{21} + V_{21} t}{Z^0_{31} + V_{31} t} \ = \ \alpha +
\frac{\beta}{
g(\zeta_3)} , \ \ \ ( \zeta_3 \ = \ \frac{\xi_{31}}{\xi_{21}} ) ,
\label{e10}\eeq
where

$$ g( \zeta ) \ = \ \frac{ \zeta^{1-\mu_1-\mu_3} {\widetilde F} ( 1- \mu_1,
\mu_2 ,
2 - \mu_1 - \mu_3 ; \zeta )}{ {\widetilde F} ( \mu_3 , \sum \mu_i - 1, \mu_1 +
\mu_3 ; \zeta )} $$
\beq \alpha \ = \ e^{i \pi \mu_1} \frac{\sin \pi \mu_3}{\sin \pi ( \mu_1 +
\mu_3 )}
 , \ \ \ \
\beta \ = \ - e^{- i \pi \mu_3} \frac{\sin \pi \mu_3 \sin \pi ( \mu_1 + \mu_2 +
\mu_3 )}{\sin \pi \mu_2 \sin \pi ( \mu_1 + \mu_3 )} \label{e11}\eeq

We thus see that $g$ represents the "mapping function" for a problem with
difference of exponents $1 - \mu_1 - \mu_3 , \ 1 - \mu_2 - \mu_3, \
\mu_1 + \mu_2 -1 $ at $\zeta_3 \ = \ 0 , 1 , \infty $ respectively.

This fact has an interesting interpretation, which is better seen in one
of the degenerate limits, for instance $\zeta_3 \ = \ \displaystyle{
\frac{\xi_{13}}{\xi_{12}}} << 1 $. This situation corresponds to close
crossing of particles \#1 and \#3
in minkowskian coordinates ( Fig. (5) ), so that the subsystem $13$ performs
an "internal" scattering. Since $\zeta_3 << 1$, and $g(\zeta) \sim
\zeta^{1- \mu_1 -\mu_3}$, Eq. (5.10) reduces in this limit to the expression

\beq {\left( \frac{\xi_{13}}{\xi_{12}} \right)}^{1 - \mu_1 - \mu_3}
\simeq \ {\rm const.} \ \frac{V_{13} t}{Z_{12}} \ , \label{e12'}\eeq
in which $\xi_{12} ( Z_{12} )$ is slowly varying with respect to $\xi_{13}
(Z_{13}) $. Therefore particles \#1 and \#3 scatter much in same way as in the
two-body case, with the static mass $m_1+m_3$ playing the role of total mass
of the subsystem. One can also verify, by using hypergeometric identities,
that Eq. (5.10) can be rewritten as

\beq \frac{Z_{13}}{Z_{12}} \ = \ \frac{g(1)}{g(1)-g(\infty)} \left( 1 -
\frac{g(\infty)}{g(\zeta_3)} \right). \  \label{e12"} \eeq

Therefore, a behaviour of type (\ref{e12'}) holds for $\zeta_3 \rightarrow 1$
( $ \xi_{23} << \xi_{12} $ ) and $\zeta_3 \rightarrow \infty$ \\ \ \
( $ \xi_{12} <<
\xi_{13} $ ) also, the relevant mass being $m_2 + m_3$ and $m_1 + m_2$
respectively. In other words, the equations for the relative shape motion,
Eq. (\ref{e10}), is clever enough to be consistent with the decoupling
properties of the two-body subsystems in the relevant limits.

A second point to be noticed is the possibility of "fixed points" of the
mapping \\
(\ref{e10}). For arbitrary initial conditions, the quantity $Z_{21} /
Z_{31}$ in the l.h.s., varying with time, describes a circle starting and
ending at $V_{21} / V_{31}$. For proper initial conditions, however, it will
be just a constant, thus yielding the equations

\beq \frac{V_{21}}{V_{31}} \ = \ \alpha + \frac{\beta}{g(\zeta_0)} \ , \
\frac{\xi_{31} (t)}{\xi_{21} (t)} \ = \ \zeta_0 \ = \ const. \label{e13}\eeq

In this situation, the $\zeta_3$ variable does not move, neither does $\eta
(\zeta_3)$. Therefore, the three particles will sit at the vertices
of a triangle,
whose angular shape is fixed by $\zeta_0$ , given implicitly by Eq.
(\ref{e13}), the
only freedom being the overall scale $\xi_{21} (t)$, which by (5.9) follows
a two-body motion

$$ {\widetilde C} {( \xi_{21} )}^{1-\sum \mu_i}  \ = \ Z^0_{21} + V_{21} t \ ,
$$
\beq {\widetilde C} / C \ = \ \int^1_0 d\zeta \ \zeta^{-\mu_1} {(\zeta -
1)}^{-\mu_2}
{(\zeta-\zeta_0)}^{-\mu_3} .  \label{e14}\eeq

Therefore, in this case there is only one "scattering angle", corresponding to
the total mass, as in Eq. (\ref{c42}). This feature also is expected to have a
generalization to arbitrary speed, an example of which will be shown in the
next subsection.

Finally, let us note that so far the explicit form of $\eta( \zeta_3)$ has
played no role for the motion. However, this feature disappears to next
order in the $V_i$'s , because the entries $O(V)$ and $O(V^2)$ in Eq.
(\ref{e8})
contain the explicit form (\ref{d16}) of $f(\zeta)$, with its $\eta(\zeta_3)$
dependence. Some further insight on the role of the $\eta$'s will appear
in the following example.

{\large \bf 5.2 A symmetric N-body case}

We have noticed that at the fixed points of Eq. (\ref{e13}), the variable
$\zeta_3 \ = \ \xi_{31}(t) / \xi_{21}(t) \ = \ \zeta_0 $ stays fixed, and the
motion is effectively of two-body type, with fixed triangular configuration
of the $\zeta$'s in the three-body case. This fact admits a generalization to
the $N$-body case by just splitting the singularities of a two-body mapping
function by a change of variables.

Consider in fact the two-body mapping function in Eq. (\ref{c24}) with $V_1
\ = \ 0$
and $V_2 \ = \ th^{-1} \eta_{21}$, and perform on it the change of variables $z
\ = \ \zeta^N$ , as follows

\beq F( \zeta ) \ = \ cth \half \eta_{21} \frac{ \zeta^{N\mu_1} {\widetilde F}
(
\half ( 1 +
\mu_{\infty} + \mu_1 - \mu_2 ), \half ( 1 - \mu_{\infty} + \mu_1 -
\mu_2 ) , 1 + \mu_1 ; \zeta^N ) }{ {\widetilde F} ( \half ( 1 + \mu_{\infty}
- \mu_1 - \mu_2 ); \half ( 1 - \mu_{\infty} - \mu_1 - \mu_2 ), 1 - \mu_1;
\zeta^N ) } . \label{e15}\eeq

The singularity at $z \ = \ 1$ is now split into $N$ ones, at $\zeta \ = \
 \omega_k \ = \
exp ( 2 \pi i k / N )$, the $N$-th roots of unity ( Fig. 6).

Correspondingly the potential, which transforms as the Schwarzian derivative,
becomes

$$ 2 Q( \zeta ) \ = \ \{ f, z \} { \left( \frac{dz}{d\zeta} \right)}^2 +
\{ z, \zeta \} \ = \ 2 q(z) {( N \zeta^{N-1} )}^2 + ( 1 - N^2 ) / 2 \zeta^2
\ = $$
\beq \ = \ \half \left[ \frac{1-{(N \mu_1 )}^2}{\zeta^2} +
\frac{ ( 1 - \mu_2^2 ) N^2 \zeta^{2N-2} }{ {( 1 - \zeta^N )}^2 } +
\frac{ N^2 ( 1 - \mu_1^2 - \mu_2^2 + \mu^2_\infty ) \zeta^{N-2} }{ 1 - \zeta^N
}
\right]. \label{e16}\eeq

As expected, this expression shows singularities at $\zeta \ = \ 0$ ($\zeta \ =
\ \omega_k $) with difference of exponents $N \mu_1 ( \mu_2 )$ respectively.
In fact, by using the identities

$$ \sum^N_{k=1} {(\zeta - \omega_k)}^{-1} \ = \ \frac{N \zeta^{N-1}}{\zeta^N
-1} ,
\sum_{k=1}^N {(\zeta - \omega_k)}^{-2} \ = \ \frac{N \zeta^{N-2} ( \zeta^N + N
-1 )
}{{(\zeta^N-1)}^2} , $$
Eq. (\ref{e16}) can be rewritten as

\beq 2 Q(\zeta) \ = \ \half \frac{( 1 - {(N\mu_1)}^2)}{\zeta^2} +
\sum^N_{n=1} \half \frac{(1-{\mu_2}^2)}{{(\zeta-\omega_k)}^2} +
\sum^N_{n=1} \frac{\beta_k}{(\zeta-\omega_k)} , \label{e17}\eeq
where

\beq \beta_k \equiv \frac{1}{2 \omega_k} [ N ( \mu_1^2 - \mu_\infty^2 ) -
( 1 - \mu_2^2 ) ] \label{e18}\eeq
are the accessory parameters.

Furthermore, by either the constraint in Eq. (\ref{d4}) or by direct inspection
of
the asymptotic behaviour in Eq. (\ref{e15}) we find

\beq \frac{{\cal M}^{(N)}}{2\pi} - 1 \ = \ \mu_\infty^{(N)} \ = \
N \mu_{\infty} \ = \ N
\left( \frac{{\cal M}^{(2)}}{2\pi} - 1 \right) \ , \label{e19}\eeq
where we have defined the total mass according to Eq. (\ref{c19}) ( Eq.
(\ref{d26}) )
for the two - \ \ ( many - ) body case. It follows that

\beq {{\cal M}}^{(N)} \ = \ N {{\cal M}}^{(2)} - 2 \pi ( N - 1 )
\label{e20}\eeq
is the total mass of the system.

Finally one can check that the monodromies at $\zeta \ =  \ \omega_k$ are those
expected for a velocity $ exp(-2\pi i \mu_1 k ) V_2 $ , and that the total
monodromy is

\beq L^{(N)} \ = \ R_1^N R^{-(N-1)}_1 L_2 R_1^{(N-1)} ...
( R_1^{-1} L_2 R_1 ) \cdot L_2 \ = \ {( R_1 L_2 )}^N , \label{e21}\eeq
corresponding to a total momentum

\beq P^{(N)} \ = \ N P^{(2)} \ , \ \ \ \ \ ( {\rm mod} 2\pi \ n \
\frac{P^{(2)}}{{\cal M}^{(2)}} ) \ , \label{e22}\eeq
$n \ = \ - ( N-1 ) $ being the choice (\ref{e20}).

Then it would seem that we have found for free an $N$-body system with masses
$N \mu_1 $ at $\zeta \ = \ 0$ , $\mu_2$ at $\zeta \ = \ \omega_k ( k \ = \
1, ..., N )$
and a nontrivial mass $\cal M^{(N)}$ at $\zeta \ = \ \infty$, the basic
simplification being the fixed symmetric positions of the singularities,
corresponding to a two-body motion of the overall scale.

A closer look shows, however, that the interpretation of the singularity at
$\zeta \ = \ 0$ should be corrected. First, notice that the total mass
(\ref{e21})
is correctly smaller than $2\pi$ if ${\cal M}^{(2)}$ is. However, its threshold
value is

\beq {\cal M}^{(N)} \ge N m_2 + N ( m_1 - 2\pi ( 1 - \frac{1}{N} )) .
\label{e23}\eeq

This suggests that the problem makes sense only if $m_1 \ge 2\pi ( 1 -
\frac{1}{N} )$, $ m_2 \le 2\pi / N $, and that the actual mass at the origin is

\beq {\cal M}^{(1)} \ = \ N m_1 - 2\pi ( N-1 )  \le 2 \pi \label{e24}\eeq
much as in Eq. (\ref{e21}). Furthermore the function $N(\zeta, t)$ for the
$N$-body
system is easily calculated from its transformation properties

\beq N^{(N)} ( \zeta ) \ = \ N^{(2)} (\zeta^N) {\left( \frac{dz}{d\zeta}
\right)}^2
\ = \ C \ {( \xi_{21} )}^{\frac{{\cal M}^{(2)}}{2\pi} - 1} \frac{N^2
\zeta^{N-2}}{
\prod_k ( \zeta - \omega_k ) } \label{e25}\eeq
and shows no pole at $\zeta \ = \ 0 $, but rather a zero of order $(N-2)$. As a
consequence

\beq \frac{N}{f'} \ \ {\buildrel {\zeta\rightarrow 0} \over \simeq} \ \
\zeta^{-
\displaystyle{\frac{N m_1}{2 \pi}} + N-1} \ \  , \ \ \ \ \label{e26}\eeq
and this is, according to the boundary condition (\ref{c26}), the behaviour
appropriate for the mass in Eq. (\ref{e23}).

Therefore the behaviour at the origin $F' \simeq \zeta^{N \mu_1 - 1}$ arises
because of the mass (\ref{e23}), degenerate with a zero of order $(N-2)$. In
particular, if $m_1$ takes the value

\beq m_1 \ = \ 2 \pi ( 1 - \frac{1}{N} ) , \ \ \ \ F' \simeq \zeta^{N-2}
\label{e27}
\eeq
there is no physical mass at the origin, but just $(N-2)$ degenerate apparent
singularities !

The above interpretation is confirmed by the explicit calculation of $\eta(
\zeta)$
in the quasi-static three-body case in Eq. (\ref{e1}). If we take equal masses,
and
we set $\zeta_3 \ = \ e^{i \pi / 3}$, so that the singularities form an
equilater
triangle, then $\eta ( e^{i \pi / 3 })$ becomes the center of such triangle,
which corresponds to the origin in the present example.

\section{Discussion}

We have given, in this paper, a clear picture of the two-body system in our
coordinates with instantaneous propagation, and we have also illustrated some
interesting  features of the classical $N$-body problem which may have an
impact on the quantized theory.

The two-body motion ( Sec. 3.3 ) is actually equivalent to a single body one,
for the relative trajectory $\xi(t) \equiv \xi_2(t) - \xi_1(t)$ which follows
the test body geodesics in the field of the total invariant mass ${\cal M}$.
Thus, the latter plays the role of hamiltonian of the system, and this
presumably justifies previous derivations of quantum mechanical amplitudes
\cite{a5},\cite{a6}, provided the classical scattering angle is correctly
identified, as in Eq. (\ref{c44}).

Some novel features appear in the many-body problem, which are related to the
"internal" motion of the dimensionless shape parameters $\zeta_i \ = \
\frac{\xi_{2i}}{\xi_{21}}$. Remarkably, the internal motion decouples, at
least in the quasi-static case, from the one of the overall scale $\xi (t)$,
which follows a two-body dynamics in the field of the total mass. The internal
dynamics is markedly different, however. In fact, it is consistent with
decoupling limits, when a two-body subsystem is singled out, but it also
admits a motion with constant shape parameters, for proper initial condition.
It is thus possible that quantization of the internal motion may yield several
surprises, including the existence of resonances in the quantum amplitudes.
This issue requires further study.

Looking now at the metric in Eq. (\ref{b29}), we see that it is expressed in
terms of Liouville fields which have various links to conformal type theories.
For instance, the Schwarzian derivative of the mapping function, which defines
particle masses and momenta in an invariant way, is related to the classical
energy-momentum tensor of such field.In fact, it is easy to show that

\beq  \{ f , z \} =  - 2 \partial_z^2{\tilde \phi}
- 2 {(\partial_z {\tilde\phi} )}^2 \ , \ \ \ {\tilde\phi} \equiv \phi -
\log |N|.  \eeq

Furthermore, the accessory parameters of the quasi-static limit in Sec. 4.2
have the same form as for vertex functions of properly defined \cite{a26}
conformal
fields, provided that the masses take over some limiting values. A similar
relationship has been advocated in Ref. \cite{a19} , on the basis of
previous work \cite{a27} on conformal parametrization of the general
Riemann-Hilbert problem, in terms of Green's functions of nonlocal vertex
operators. Although such remarks do not help much making the solution explicit,
they show that more attention should be devoted to direct
quantization of the longitudinal degrees of freedom of the problem \cite{a28},
coupled to matter fields.

We should also point out some left over problems, even at the
classical level.

We have not treated at all spinning particles, because in this case the
Minkowskian coordinates are not well defined at the particle sites, due to
the known time shift \cite{a7} in their rest frame. This suggests that the
mapping
to single-valued coordinates becomes singular close to the particles. In fact
the ${\delta}'$ singularity in the energy momentum tensor representing the
localized spin, implies a double pole in the meromorphic $N$-function, and
then a singularity of the mapping and a Gribov horizon ( $|f| = 1$ ). Thus,
a more refined analysis is needed.

The results of this paper in the instantaneous gauge do not help
understanding the issue of matter asymptotic states in (2+1)-gravity, if any.
In fact, there is no way of decoupling particles for large times in the
two-body case, and only in a limited way for the limit of $N+1$ bodies to $N$.
This is to be contrasted to the yet partial results obtained \cite{a10} in
covariant gauges, where such decoupling is possible, but a redefinition of
energy \cite{a21} is needed, and thus of scattering parameters ( Sec. 3.5 ).
The trouble is that the definition of the localized energy-momentum and of the
scattering matrix is dependent on the coordinate frame and thus on the gauge
choice.

On the whole, we feel however that the present treatment of the $N$-body
problem
has helped clarifying a lot the degrees of freedom of 2+1 Gravity with matter
and thus may be the basis of further progress in the quantum theory much
as already occured in conformal models.

{\bf Acknowledgements }

It is a pleasure to thank Luis Alvarez-Gaum\`e, Andrea Cappelli, Camillo
Imbimbo, Giorgio Longhi, Pietro Menotti, Gabriele Veneziano and Erik Verlinde
for interesting discussions. One of us (M.C.) is grateful
to the CERN Theory Division for hospitality, while part of this work was
being done.

\appendix
\section{Appendix - Integrals for the two-body motion}

We show in the following the derivation of the relevant integrals $I^0$, $I^z$,
$I^{\bar z}$ necessary to make explicit the geodesic equation (\ref{b35}) for
the two-body problem.

The three integrals have the following structure

\beq I^a (0, \zeta )
\ = \ \int^\xi_0 \ \frac{dz}{z(1-z)} \ y_\alpha y_\beta \ c^a_{\alpha
\beta}, \label{ap1} \eeq
where the only non-vanishing entries for the coefficients $c^a_{\alpha\beta}$
are
\beq c^0_{+-} = 1, \ \ c^z_{--} = 1, \ \ c^{\bar z}_{++} = 1. \label{ap2} \eeq

The integrals $I^a$ can be computed exactly by deriving Eq (\ref{c7}) with
respect to $\mu^2_\infty$ to obtain the special measure $\frac{1}{z(1-z)}$
and therefore the following identity:

\beq \frac{y_\alpha y_\beta}{z(1-z)} \ = \ \frac{d}{dz} W \left( y_\alpha,
\frac{dy_\beta}{d\mu^2_\infty} \right) \label{ap3} \eeq

Specializing Eq. (\ref{ap1}) to the case of $f_{(1)}$ in Eq. (\ref{c24}),
we obtain

$$ I^0 ( 0, \xi )
= \int^\xi_0 \frac{dz}{z(1-z)} y_{+} y_{-} = + \frac{2}{\mu_\infty} {\left(
\log F( a, b, c, \xi ) \right) }_{/ \mu_\infty} \ -  $$ $$ - 2
\frac{k_{+} k_{-}}{\mu_\infty} \ \xi \ {\tilde F} ( a, b, c, \xi )
{\tilde F} ( a', b', c', \xi ) {\left( \frac{ab}{c} \frac{ F( c-a, c-b, c+1 ;
\xi )}{F( a, b, c ; \xi )} \right)}_{/ \mu_\infty} , $$

$$ I^z (0,\xi)
\ = \ \int^\xi_0 \frac{dz}{z(1-z)} y_{-}^2 \ = \  - 2 \
\frac{k_{-}^2}{\mu_\infty} \ \xi^{2\lambda^{-}_1} \ {\tilde F}^2 ( a, b,
c, \xi ) {\left( \frac{ab}{c} \frac{ F( c-a, c-b, c+1 ;
\xi )}{F( a, b, c ; \xi )} \right)}_{/ \mu_\infty} \ , $$

\beq  I^{\bar z} (0, \xi )
\ = \ \int^\xi_0 \frac{dz}{z(1-z)} y_{+}^2 \ = \ - 2 \
\frac{k_{+}^2}{\mu_\infty} \ \xi^{2\lambda^{+}_1} \ {\tilde F}^2 ( a',
b', c' , \xi ) {\left( \frac{ab}{c} \frac{ F( c'-a', c'-b', c'+1 ;
\xi )}{F( a, b, c ; \xi )} \right)}_{/ \mu_\infty} \label{ap4} \eeq

where

$$ k^2_{-} \ = \ \frac{\gamma_{12} V_{21} }{\gamma_{12} + 1}
\frac{\pi}{\sin \pi \mu_1}
\frac{1}{\Gamma (a) \Gamma (b) \Gamma (a') \Gamma (b')} \ ,  $$
\beq k^2_{+} \ = \ \frac{\gamma_{12} {\bar V}_{21} }{\gamma_{12}
 - 1} \frac{\pi}{\sin \pi
\mu_1} \frac{1}{\Gamma (a) \Gamma (b) \Gamma (a') \Gamma (b')}. \label{ap5}
\eeq

By then setting $\xi= 1$ we obtain the final result
$$ I^0_{(1)}
\ = \ \frac{1}{\mu_\infty} \frac{\sin \pi a \sin \pi b}{ \sin \pi \mu_1
\sin \pi \mu_2} [ \psi (a) - \psi (b) ] +
\frac{1}{\mu_\infty} \frac{\sin \pi a' \sin \pi b' }{ \sin \pi \mu_1
\sin \pi \mu_2} [ \psi (1-a') - \psi (1-b') ] \ , $$
$$ I^z_{(1)}
\ = \ \frac{1}{\mu_\infty} \frac{\gamma_{12} V_{21} }{\gamma_{12} + 1}
\frac{\sin \pi a' \sin \pi b' }{ \sin \pi \mu_1
\sin \pi \mu_2} [ \psi (a) - \psi (b) + \psi (1-a') - \psi (1-b') ] \ , $$
\beq I^{\bar z}_{(1)}
\ = \  \frac{1}{\mu_\infty} \frac{\gamma_{12} {\bar V}_{21} }{\gamma_{12} - 1}
\frac{\sin \pi a \sin \pi b }{ \sin \pi \mu_1
\sin \pi \mu_2} [ \psi (a') - \psi (b') + \psi (1-a) - \psi (1-b) ] \ ,
\label{ap6} \eeq
where the subscript $(1)$ refers to the rest frame of particle \#1, and
$\psi (z) \ = \ \frac{d \log \Gamma (z)}{dz} $.  Since $\psi (z) - \psi (1-z)
\ = \ \pi \cot \pi z $, we derive from (\ref{ap6}) the simpler formula
(\ref{c46}) used in the text.

Let us comment about the vector property of $I^a(0,\xi)$. First, note that
$I_{(1)}^a ( 0, \zeta )$ changes by just a phase when $z$ turns around $\xi_1$,
this being the relevant monodromy. So does $I^a_{(2)}$ when $z$ turns around
$\xi_2$. But $f_{(2)}$ and $f_{(1)}$ are related by a boost to the rest frame
of particle \# 2. Therefore $I_{(1)}^a$ has the correct monodromy around each
particle, and the mapping in Eq. (\ref{c33}) automatically satisfies the
complete DJH matching conditions in Eq. (2.10), including the translational
part.



\newpage

\newpage
\begin{center}
{\bf Figure Captions}
\end{center}
\vspace{1cm}

{\bf Fig.1}:   Particles with tails, and related monodromies.
\vspace{.7cm}

{\bf Fig.2}:   a) Schwarz triangle, for the mapping function $f_{(1)}(z) $
         in the case
         ${\cal M} < 2\pi $, where $ f_{\_ }(\infty )$ is the Schwarz
         reflection of
         $f_{+}\left( \infty \right)$ through the segment $[f(o), f(1)]$.
         b) The same for ${\cal M}=2\pi \left( 1 + i\sigma \right)$; the
         intermediate closed curve is image of a straight line with constant
         $\theta = arg(\zeta )$. The critical
         value $f\bar f = 1 $ is crossed an infinity of times.
\vspace{.7cm}

{\bf Fig.3}:   Tails and values of $f_(\infty)$ for a given cyclic ordering of
         N particles.
\vspace{.7cm}

{\bf Fig.4}:   f - Mapping for $ N=3 $. The deficit angles at particle sites
         are shown
         together with the ones at space infinity, whose sum is
         $2\pi - {\cal M}$.
\vspace{.7cm}

{\bf Fig.5}:   Scattering of a two-body subsystem in the three-body motion.
\vspace{.7cm}

{\bf Fig.6}:   Configuration of tails for the symmetric N-body problem.
\vspace{.7cm}
\end{document}